%% file: g301.tex
\documentclass[aas_macros,a4paper,twoside]{article}
\newcommand{\affil}[1]{$^{\rm #1}$}
\usepackage[authoryear]{natbib}
\bibpunct{(}{)}{;}{a}{}{,}
\usepackage{aas_macros}
\usepackage{graphicx}
\usepackage{fix2col}
\usepackage[lofdepth,lotdepth]{subfig}
\usepackage{longtable}
\usepackage[tableposition=t]{caption}
\usepackage[breaklinks,colorlinks,citecolor=blue]{hyperref}
\date{} 
\newcommand{\kms}{\mbox{km\,s$^{-1}$}}
\newcommand\arcdeg{\mbox{$^\circ$}}%
\newcommand\arcmin{\mbox{$^\prime$}}%
\newcommand\arcsec{\mbox{$^{\prime\prime}$}}%

\newcommand{\nthp}{N\ensuremath{_2}H\ensuremath{^+}}

\newcommand{\tcts}{$^{13}$C$^{34}$S}
\newcommand{\tcs}{$^{13}$CS}
\newcommand{\cth}{C$_{2}$H}
\newcommand{\chtcn}{CH$_{3}$CN}

\newcommand{\htcop}{H$^{13}$CO$^{+}$}

\newcommand{\hctn}{HC$_{3}$N}
\newcommand{\hntc}{HN$^{13}$C}

\newcommand{\hctccn}{HC$^{13}$CCN}
\newcommand{\hcn}{HCN}
\newcommand{\hnc}{HNC}
\newcommand{\hcop}{HCO\ensuremath{^{+}}}

\newcommand{\sio}{SiO}

\newcommand{\UCHII}{UC~H\scriptsize{II}~\small{region}}
\newcommand{\HII}{H\scriptsize{II}~\small{region}}

\newcommand{\Tsys}{\ensuremath{T_{\rm sys}}}

\title{\large\bf\flushleft Characterization of the MALT90 Survey and the Mopra Telescope at 90 GHz}
\author{\parbox{\textwidth}{\flushleft
\vspace{-0.5cm}
{\it Foster, J.B.\affil{A,B,R}, Rathborne, J.M.\affil{C}, Sanhueza, P.\affil{B}, Claysmith, C.\affil{B}, Whitaker, J.S.\affil{D}, Jackson, J.M.\affil{B}, Mascoop, J.L.\affil{B}, Wienen, M.\affil{E}, Breen, S.L.\affil{C}, Herpin, F.\affil{F,G}, Duarte-Cabral, A.\affil{F,G}, Csengeri, T.\affil{E}, Contreras, Y.\affil{C}, Indermuehle, B.\affil{C}, Barnes, P.J.\affil{H}, Walsh, A.J.\affil{I}, Cunningham, M.R.\affil{J}, Britton, T.R.\affil{C,K}, Voronkov, M.A.\affil{C}, Urquhart, J.S.\affil{E}, Alves, J.\affil{L}, Jordan, C.H.\affil{M,C}, Hill, T.\affil{N,O}, Hoq, S.\affil{B}, Brooks, K.J.\affil{C}, Longmore, S.N.\affil{P,Q}}\\
\vspace{0.4cm}
{\small \affil{A}\, Yale Center for Astronomy and Astrophysics, Yale University, New Haven, CT 06520, USA}\\
{\small \affil{B}\, Institute for Astrophysical Research, Boston University, Boston, MA 02215, USA}\\
{\small \affil{C}\, CSIRO Astronomy and Space Science, PO Box 76, Epping, NSW, 1710, Australia}\\
{\small \affil{D}\, Physics Department, Boston University, Boston, MA 02215, USA}\\
{\small \affil{E}\, Max-Planck-Institut f\"{u}r Radioastronomie, Auf dem H\"{u}gel 69, D-53121 Bonn, Germany}\\
{\small \affil{F}\, Univ. Bordeaux, LAB, UMR 5804, F-33270, Floirac, France}\\
{\small \affil{G}\, CNRS, LAB, UMR 5804, F-33270, Floirac, France}\\
{\small \affil{H}\, Astronomy Department, University of Florida, Gainesville, FL 32611, USA}\\
{\small \affil{I}\, International Centre for Radio Astronomy Research, Curtin University, GPO Box U1987, Perth WA 6845, Australia}\\
{\small \affil{J}\, School of Physics, University of New South Wales, Sydney, NSW, 2052, Australia}\\
{\small \affil{K}\, Dept. of Physics \& Astronomy, Macquarie University, Sydney, NSW, 2109, Australia}\\
{\small \affil{L}\, Department of Astrophysics, University of Vienna, T\"{u}rkenschanzstrasse 17, 1180, Vienna, Austria}\\
{\small \affil{M}\, School of Mathematics and Physics, University of Tasmania, Private Bag 37, Hobart, Tasmania 7001, Australia}\\
{\small \affil{N}\, Laboratoire AIM Paris-Saclay, CEA/IRFU-CNRS/INSU-Universit\'{e} Paris Diderot, CEA Saclay, 91191 Gif-sur-Yvette Cedex, France}\\
{\small \affil{O}\, Joint ALMA Observatory, Alonso de C\'{o}rdova 3107, Vitacura 763-0355, Santiago, Chile}\\
{\small \affil{P}\, Astrophysics Research Institute, Liverpool John Moores University, Twelve Quays House, Egerton Wharf, Birkenhead CH41 1LD, UK}\\
{\small \affil{Q}\, European Southern Observatory, Karl-Schwarzschild-Strasse. 2, D-85748 Garching bei M\"{u}nchen, Germany}\\
{\small \affil{R}\, E-mail: jonathan.b.foster@yale.edu}}}
\begin{document}
\twocolumn[
\begin{changemargin}{.8cm}{.5cm}
\begin{minipage}{.9\textwidth}
\vspace{-1cm}
\maketitle
%
%
\small{\bf We characterize the Millimeter Astronomy Legacy Team 90 GHz (MALT90) Survey and the Mopra telescope at 90 GHz. We combine repeated position-switched observations of the source G300.968$+$01.145 with a map of the same source in order to estimate the pointing reliability of the position-switched observations and, by extension, the MALT90 survey; we estimate our pointing uncertainty to be 8\arcsec. We model the two strongest sources of systematic gain variability as functions of elevation and time-of-day and quantify the remaining absolute flux uncertainty. Corrections based on these two variables reduce the scatter in repeated observations from 12-25\% down to 10-17\%. We find no evidence for intrinsic source variability in G300.968$+$01.145. For certain applications, the corrections described herein will be integral for improving the absolute flux calibration of MALT90 maps and other observations using the Mopra telescope at 90 GHz.}


\medskip
\medskip
\end{minipage}
\end{changemargin}
]
\small

\section{Introduction}

The Millimeter Astronomy Legacy Team 90 GHz Survey (MALT90) is characterizing the physical and chemical conditions of dense molecular clumps associated with high-mass star formation over a wide range of evolutionary states using the ATNF (Australia Telescope National Facility) Mopra 22-m radiotelescope \citetext{Jackson et al. in prep.}. MALT90 targets are chosen from the APEX (Atacama Pathfinder Experiment) Telescope Large Area Survey of the Galaxy \citep[ATLASGAL;][]{Schuller:2009, Contreras:2013}. This paper presents an analysis of G300.968$+$01.145 (G301), and uses our repeated position-switched (PSW) observations of this source to characterize the system performance of the Mopra telescope as used in the MALT90 survey, including the pointing reliability, systematic gain variation, and the absolute flux uncertainty. 

Our primary goal in observing G301 was to ascertain good system performance at the start of each observing session. G301 was chosen based on its Galactic position and its relative brightness in many transitions in the MALT90 pilot survey \citep{Foster:2011}. A typical observing session commenced with a pointing on an SiO maser, followed by a PSW observation of G301 and a quick examination of the resulting spectrum. If the transitions of G301 were detected at the expected level, it demonstrated that the system was working well, and the rest of the observations continued. 

Although this system check was the main purpose of the PSW observations of G301, our repeated observations of the same source under different conditions allows us to carry out a detailed assessment of the systematics and error budget of our survey, as well as to characterize aspects of the performance of the Mopra telescope at 90 GHz. The beam size, beam shape, and efficiency of the Mopra telescope have previously been measured at 90 GHz and 16-50 GHz \citep{Ladd:2005, Urquhart:2010}, and this paper focuses on characterizing other aspects of the Mopra telescope at 90 GHz.

G301 is a molecular clump associated with the ultracompact \HII\ (\UCHII) known as RCW~65 \citep{Rodgers:1960} or Gum~43 \citep{Gum:1955}. G301 contains prominent OH masers at 1665 MHz and 1667 MHz; it has been studied extensively over the past 40 years \citep[e.g.][]{Robinson:1974} and has been found to contain numerous other maser features including OH masers at 6035 and 6030 MHz \citep{Caswell:2009} and a methanol maser at 6668 MHz \citep{Caswell:1997}. On the basis of the maser data, \citet{Caswell:2009} conclude that this source is a canonical example of an OH maser in a high-mass star-forming region, with a cluster of maser spots projected against an \UCHII; they regard the source as near the end of the evolutionary period in which it is capable of supporting maser emission, suggesting an age for the \UCHII\ near the lifetime of such objects, $\sim 10^5$ years \citep{Churchwell:1999}. 

As a well-known southern high-mass star-forming region, G301 has been included in a large number of studies, including searches for other masers \citep[e.g.][]{Caswell:2003, Dodson:2002}, studies measuring the magnetic fields in \HII s \citep[e.g][]{Han:2007}, continuum surveys of southern regions of high-mass star formation \citep[e.g][]{Faundez:2004,Walsh:2001,Walsh:1999,Walsh:1998,Kwok:1997}, and observations of dense gas tracers such as NH$_3$ (1,1) \citep{Vilas-Boas:2000}, C$^{18}$O (2-1) and HNCO (10$_{0,10}$ - 9$_{0,9}$) \citep{Zinchenko:2000}, and isotopologues of CS \citep{Chin:1996}. The 6-GHz Methanol Multibeam (MMB) survey used G301 to check their calibration stability \citep{Green:2009}.

\UCHII s have variable continuum emission on the timescale of years \citep[e.g][]{Franco-Hernandez:2004, Galvan-Madrid:2008}. It is possible, therefore, that the molecular line emission from an \UCHII\ such as G301 could also be variable on these timescales. Typical timescales for significant changes in molecular abundances due to chemistry are $>10^3$ years \citep[e.g.][]{vanDishoeck:1998, Viti:2004} although some chemistry in ``hot cores'' around massive protostars may take place on timescales of $10^{2.5}$ years \citep[][]{ChapmanJF:2009}. This is still long compared to the timescale for continuum variability. 

In the simulations of \citet{Peters:2010} and \citet{Galvan-Madrid:2011}, the continuum variability in an \UCHII\ arises from the shielding of the ionizing source by its own accretion flow. They note that since the mass of ionized gas is typically much less than the mass of molecular gas observed in an \UCHII, the variability of the molecular gas due to small clumps of mass becoming ionized or recombining would be much less than the continuum variability. Another model which explains \HII\ region variability via variations in the ionizing source itself \citep{Klassen:2012} does not produce sufficiently large continuum emission variability on the appropriate time scales to account for the observations of \citet{Franco-Hernandez:2004} and \citet{Galvan-Madrid:2008}. In the \citet{Klassen:2012} model, molecular line emission should vary only on timescales of thousands of years. Therefore we consider it highly unlikely that the molecular line emission from G301 will be intrinsically variable, although we briefly consider this possibility. 

\section{Observations}
\label{sec:observations}

\subsection{Position Switched Observations}
\label{sec:PSWobs}
We typically observed G301 once at the beginning of each observing session. Throughout this paper we shall refer to a single block of observing time as a ``session'' or an ``observing session'' (sessions were typically 11 - 14 hours in duration) and we shall use the term ``season'' or ``observing season'' to refer to the time period during which our observations were conducted during the year. We had three observing seasons from July to September in 2010, from May to October in 2011, and from May to October in 2012. During the first two observing seasons, G301 was typically between 35\arcdeg - 40\arcdeg\ of elevation at the start of our observing sessions. During our third observing season, sessions started at a later local sidereal time, so G301 was typically between 55\arcdeg - 60\arcdeg\ of elevation at the start of an observing session. 

We occasionally obtained additional observations of G301 during a given session or at atypical elevations for a variety of reasons, such as (1) mechanical failure or bad weather delaying the start of an observing session, (2) a non-standard start time for an observing session (due to the schedule of other projects), and (3) in order to better characterize G301 for this analysis. We performed a total of 258 observations of G301. A small number (10) of observing sessions for MALT90 started after G301 had set, in which case system checks were performed on another source (G337.005$+$00.323), but the sample of observations of G337.005$+$00.323 is too small to be useful for characterization and is not considered here. 

We obtained a single PSW observation with 150 seconds of on source integration time interlaced with an equal amount of time spent on a reference position at +1\arcdeg\ in Galactic latitude. The observing pattern was OFF-ON-ON-OFF-OFF-ON-ON-OFF-OFF-ON with individual integrations of 30 seconds. Both linear polarizations were observed, and were averaged together for all the following analysis. 

A PSW observation of G301 always immediately followed a successful pointing correction routine on an SiO maser (hereafter we refer to this process as ``pointing''). In subsequent observing, we pointed on a SiO maser before every source, roughly once an hour. The pointing precision of PSW observations G301, immediately following a pointing correction, is therefore typical of the pointing precision of our maps. Several different SiO masers were used as the pointing source for PSW observations of G301. During the first observing season we most commonly used X Cen, and during the second and third observing seasons we most commonly used RW Vel. In addition, we sometimes used IRSV 1540, W Hya and VX Sgr\footnote{See \url{http://www.narrabri.atnf.csiro.au/cgi-bin/obstools/siomaserdb.cgi} for details of these SiO masers}. Unfortunately, the often strong intrinsic brightness variability of SiO masers and our inconsistent use of a single pointing source precludes us from being able to use the brightness of the pointing source for characterization. 

Immediately following a PSW observation of G301, we returned to perform a pointing correction on an SiO maser (the particular maser varied based on the location of the source to be subsequently observed). The offsets (in azimuth and elevation) deduced from this pointing correction routine were recorded automatically and these offsets can be used as an additional estimate of pointing precision. 

We observed G301 using the same frequency setup as for the full survey \citetext{Rathborne et al. in prep.}, with 16 spectral windows of 138 MHz each providing $\sim0.11$ \kms\ velocity resolution around 16 rest frequencies corresponding to our targeted transitions. In this paper we focus on the four strongest transitions, highlighted in bold in Table~\ref{table:if_table}. These are all ground-state ($J = 1 - 0$) transitions, and henceforth we shall refer to these transitions only by the molecule or ion (i.e. \nthp\ instead of \nthp\ J = 1-0).

\begin{table}[thbp]
\small
\begin{center}
\caption{Spectrometer Configuration}
\label{table:if_table}
\begin{tabular}{llll}
\hline IF$^a$ & Species & Main Transition & $\nu$(GHz)$^b$\\
\hline 
\textbf{0}	&	\textbf{\nthp} &       \boldmath{$J = 1 - 0$}                                              & \textbf{93.17377}\\
1	&	\tcs		&	$J = 2 - 1$         						& 92.49430\\
2	&	H		&	41$\alpha$       						& 92.03448\\
3	&	\chtcn	&	$J_K = 5_1 - 4_1$					& 90.97902\\
4	&	\hctn		&	$J = 10 - 9 $						& 91.19980\\
5	&	\tcts		&	$J = 2 - 1$              					& 90.92604\\
\textbf{6}	&	\textbf{\hnc}		&	\boldmath{$J = 1 - 0$}						& \textbf{90.66357}\\
7	&	\hctccn	&	$J = 10 - 9, F = 9-8$					& 90.59306\\
\textbf{8}	&	\textbf{\hcop}	&	\boldmath{$J = 1 - 0$}						& \textbf{89.18853}\\
\textbf{9}	&	\textbf{\hcn}		&	\boldmath{$J = 1 - 0$}						& \textbf{88.63185}\\
10	&	HNCO	&	$J_{K_a,K_b}=4_{0,4}-3_{0,3}$		& 88.23903\\
11	&	HNCO	&	$J_{K_a,K_b}=4_{1,3}-3_{1,2}$		& 87.92524\\
12	&	\cth		&	$N = 1 - 0$						& 87.31692\\
	&			&	$J = \frac{3}{2} - \frac{1}{2}, F =  2-1$		&		    \\
13	&	\hntc		&	$J = 1 - 0$						& 87.09086\\
14	&	\sio		&	$J = 2 - 1$						& 86.84701\\
15	&	\htcop	&	$J = 1 - 0$						& 86.75433\\
\hline
\multicolumn{4}{p{7cm}}{{$^a$ This paper will focus on the four transitions shown in bold in this table.}}\\
\multicolumn{4}{p{7cm}}{{$^b$ Uncertainties on rest frequencies are less than the spectral resolution.}}\\

\medskip\\
\end{tabular}

\end{center}

\end{table}

\subsection{Mapping Observations}

In addition to the PSW observations, we obtained three 3\arcmin x 3\arcmin\ maps of G301. The first map was obtained as part of the regular survey, and subsequent maps were obtained to increase the signal-to-noise-ratio of the map and to assist in measuring the pointing and flux uncertainty. These maps were taken in the normal mode for the survey \citetext{see Rathborne et al. in prep.}, with two on-the-fly maps made by scanning both in Galactic latitude and Galactic longitude. We consider only the \Tsys\ weighted co-addition of the two different scan maps in this analysis since the pointing error between the two scan maps made in different directions will be minimal. Table~\ref{table:maps} displays the UT date and time, as well as the measured \Tsys\ and elevation, of the maps of G301.

\section{Reduction}
\label{sec:reduction}

Reduction of the PSW observations were carried out in the \textsc{asap}\footnote{\url{http://svn.atnf.csiro.au/trac/asap}} package by (1) producing a quotient spectrum from adjacent ON and OFF observations, (2) performing frequency alignment (of minimal importance during such a short series of observations), (3) averaging the two linear polarizations together using \Tsys\ weighting, and (4) averaging the five different ON-OFF cycles using \Tsys\ weighting. Finally, we fit the baseline within each IF with a second order polynomial, excluding 300 channels (out of a total of 4096) at the edge of each IF. Note that this procedure does not include a gain-elevation correction, as this has not been accurately measured for the Mopra telescope at 90 GHz; the derivation of the gain-elevation correction from these data is one of the goals of this paper. Because the Mopra telescope uses a paddle for \Tsys\ calibration at 90 GHz, our data is already opacity corrected. 

\begin{table}[t]
\small
\begin{center}
\caption{Maps of G301}\label{table:maps}
\begin{tabular}{llcc}
\hline UT Date\_Time & Dir. & $<$\Tsys$>$ & $<$Elevation$>$\\
				&			&	[K]		& [Deg.]	\\
\hline 
2011-05-06\_1527	&	GLat		&	169.0	&	41.5 \\
2011-05-06\_1558	&	GLon	&	173.5	&	37.8 \\
2011-08-22\_0034	&	GLat		&	160.5	&	42.4 \\
2011-08-22\_0104	&	GLon	&	159.8	&	46.0 \\
2012-06-29\_0657	&	GLat		&	184.1	&	58.6 \\
2012-06-29\_0727	&	GLon	&	181.0	&	59.6 \\
\hline
\end{tabular}
\medskip\\
\end{center}
\end{table}

The maps of G301 were reduced using the MALT90 reduction pipeline, which uses the ATNF packages \textsc{Livedata} and \textsc{Gridzilla}\footnote{\url{http://www.atnf.csiro.au/computing/software/livedata/index.html}} to produce a map from the on-the-fly data. The pipeline performs reference subtraction (with reference positions $\pm$1$\arcdeg$ away from the Galactic plane), polarization averaging, baseline subtraction with a second order polynomial fit (excluding 300 channels on the edge of the bandpass out of a total of 4096 channels) and \Tsys\ weighted co-addition of the spectra within the maps to produce a lightly smoothed map with an effective beam of 38\arcsec. Our modified pipeline version of {\sc Livedata} applies an 11-channel Hanning smoothing kernel to the reference spectra before subtracting them from the source spectra in order to mitigate striping artifacts in the maps.

All data in this paper is presented on the antenna temperature $T_A^*$ scale. The main beam efficiency for the Mopra telescope at 90 GHz was estimated to be 0.49$\pm$0.03 by \citet{Ladd:2005}. For compact sources ($<$80\arcsec), division by this number would approximately convert our antenna temperature measurements into main-beam brightness temperatures, although additional efficiency corrections (i.e. gain factors) are derived in this work which suggests that additional corrections are required. 

\subsection{Spectral Line-Fitting}

\begin{figure*}[htbp]
\subfloat[][\nthp]{\includegraphics[width=0.48\linewidth]{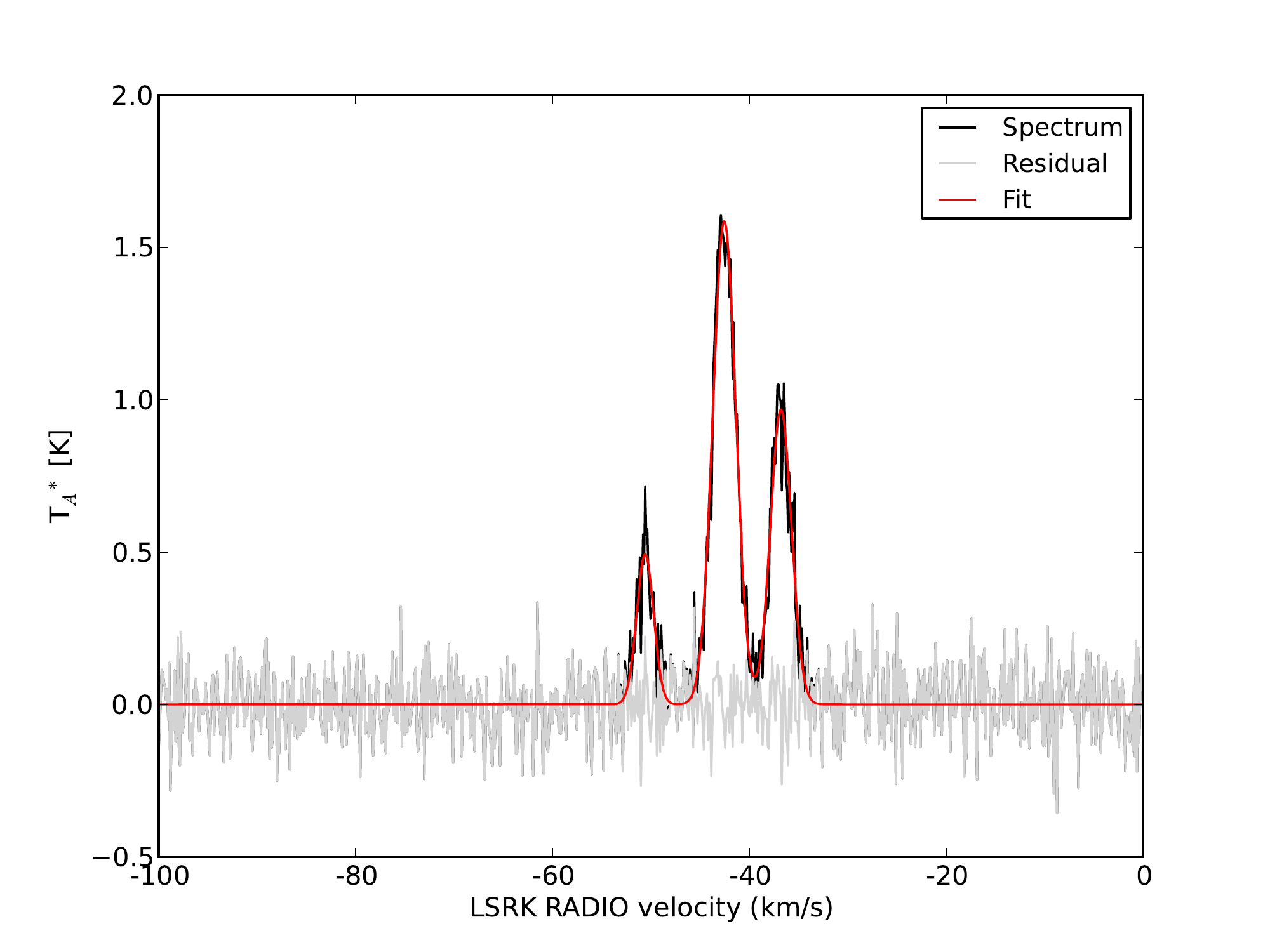}
\label{fig:examplefits_a}}
\qquad
\subfloat[][\hnc]{\includegraphics[width=0.48\linewidth]{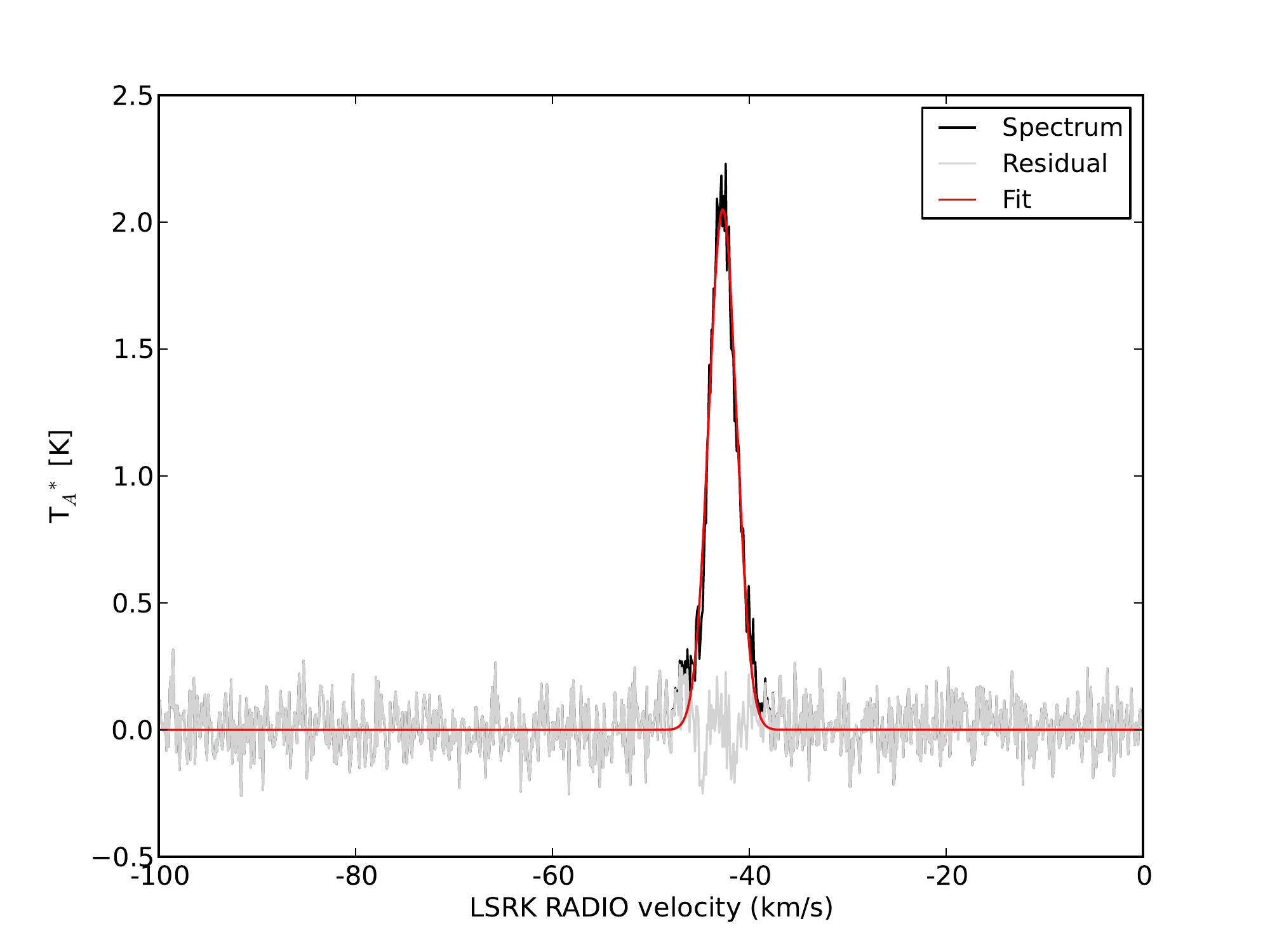}
\label{fig:examplefits_b}}
\qquad
\subfloat[][\hcop]{\includegraphics[width=0.48\linewidth]{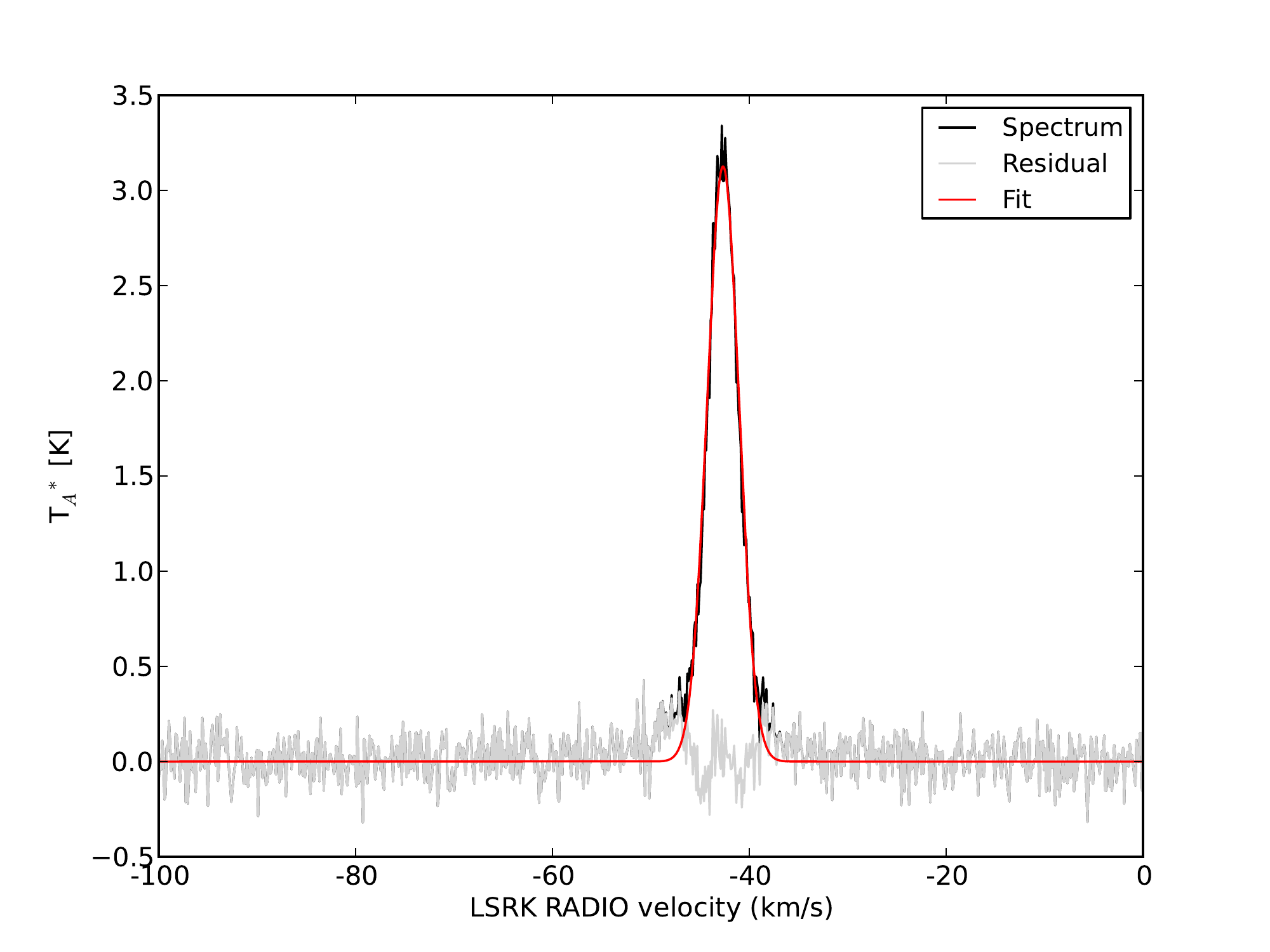}
\label{fig:examplefits_c}}
\qquad
\subfloat[][\hcn]{\includegraphics[width=0.48\linewidth]{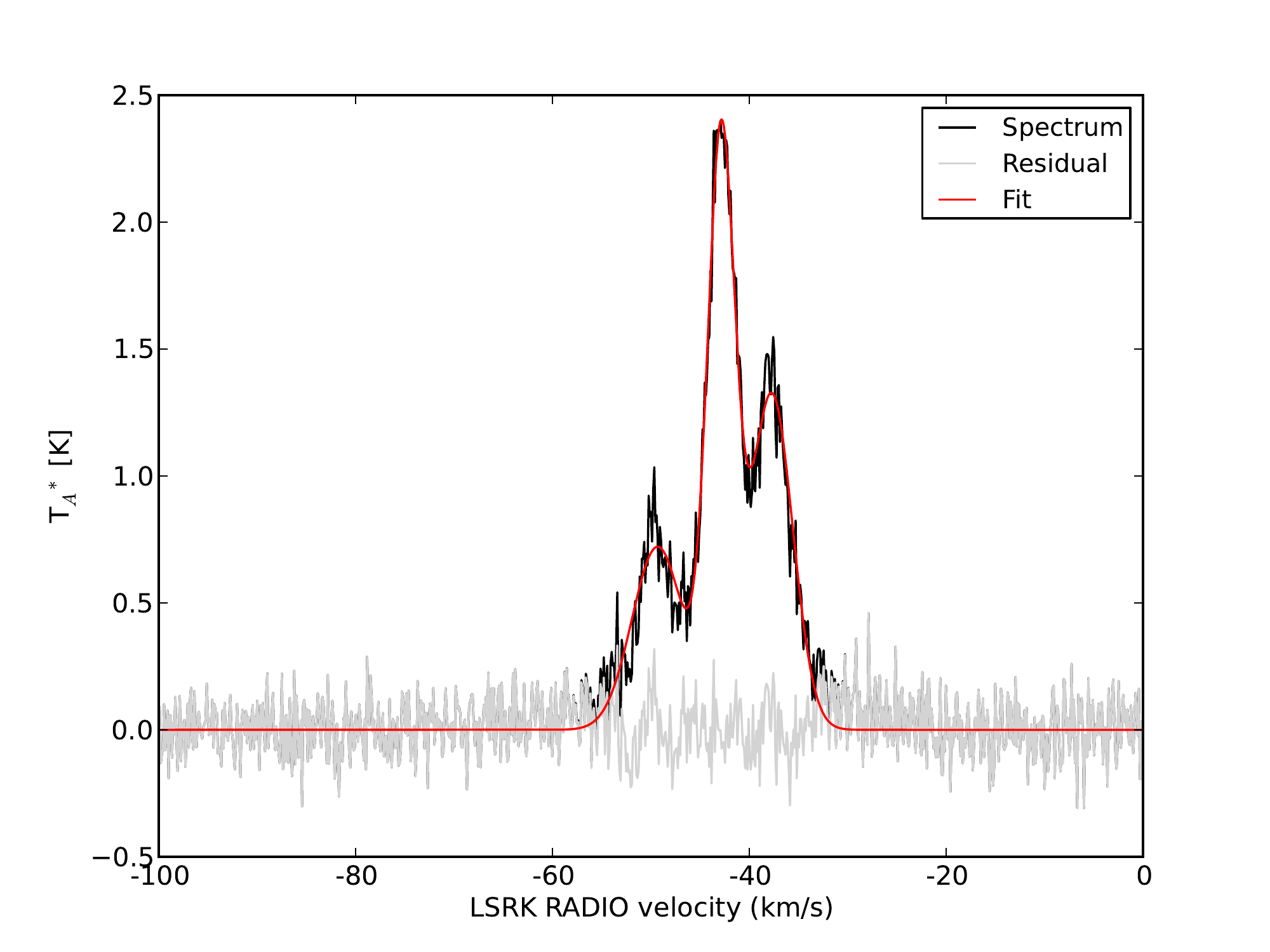}
\label{fig:examplefits_d}}
\caption{Gaussian fits of the four strongest transitions for a typical PSW spectrum of G301 (2011-09-26\_1). The data are in black, the fit is shown in red, and the residual is in gray. }
\label{fig:examplefits}
\end{figure*}

Following basic reduction, we fit the four strongest transitions (\nthp, \hnc, \hcop, and \hcn) with a number of Gaussians corresponding to the number of resolved components present. \nthp\ and \hcn\ are each fit with three Gaussians with fixed velocity separations and initial intensity ratios appropriate for the optically thin hyperfine components. \hnc\ and \hcop\ are fit with single Gaussians. \textsc{asap} estimates the Gaussian parameters and associated uncertainty from the noise in the spectra. Fitting results for a typical PSW spectrum toward G301 are shown in Figure~\ref{fig:examplefits}. This observation is typical in the sense that it is the closest to the median in \Tsys\ (179 K) and elevation (50.37\arcdeg). 

Spectra in the maps were fit using \textsc{specfit} within \textsc{casa}\footnote{\url{http://casa.nrao.edu/}}, with the same parameters as for the PSW observations. \textsc{specfit} produces output maps of the fit parameters and automatically masks pixels within the map that fail to produce a reliable fit. \hnc\ and \hcop\ are reliably fit over most of the map, while \nthp\ and \hcn\ are only reliably fit over the central portion of the map.

\section{Data Description}
\label{sec:data}
\subsection{Position Switched Data}
\label{sec:pointeddata}

Tables describing the PSW observations of G301 and our Gaussian fit parameters are given in the Appendix. In the rest of the analysis we consider only observations for which the fit parameters were well determined according to the following criteria. For a given molecular transition, $n$, we require that the amplitude, $a$ and the fit uncertainties on the amplitude and velocity ($\sigma_{a_n}$ and $\sigma_{v_n}$) obey the following:

\begin{equation}
a_n > 0 
\label{eqn:reasonable-start}
\end{equation}
\begin{equation}
\sigma_{a_n} < 0.2 K
\end{equation}
\begin{equation}
\sigma_{v_n} < 0.07 \kms
\label{eqn:reasonable-end}
\end{equation}

\begin{figure}[th]
\begin{center}
\includegraphics[width=\linewidth, angle=0]{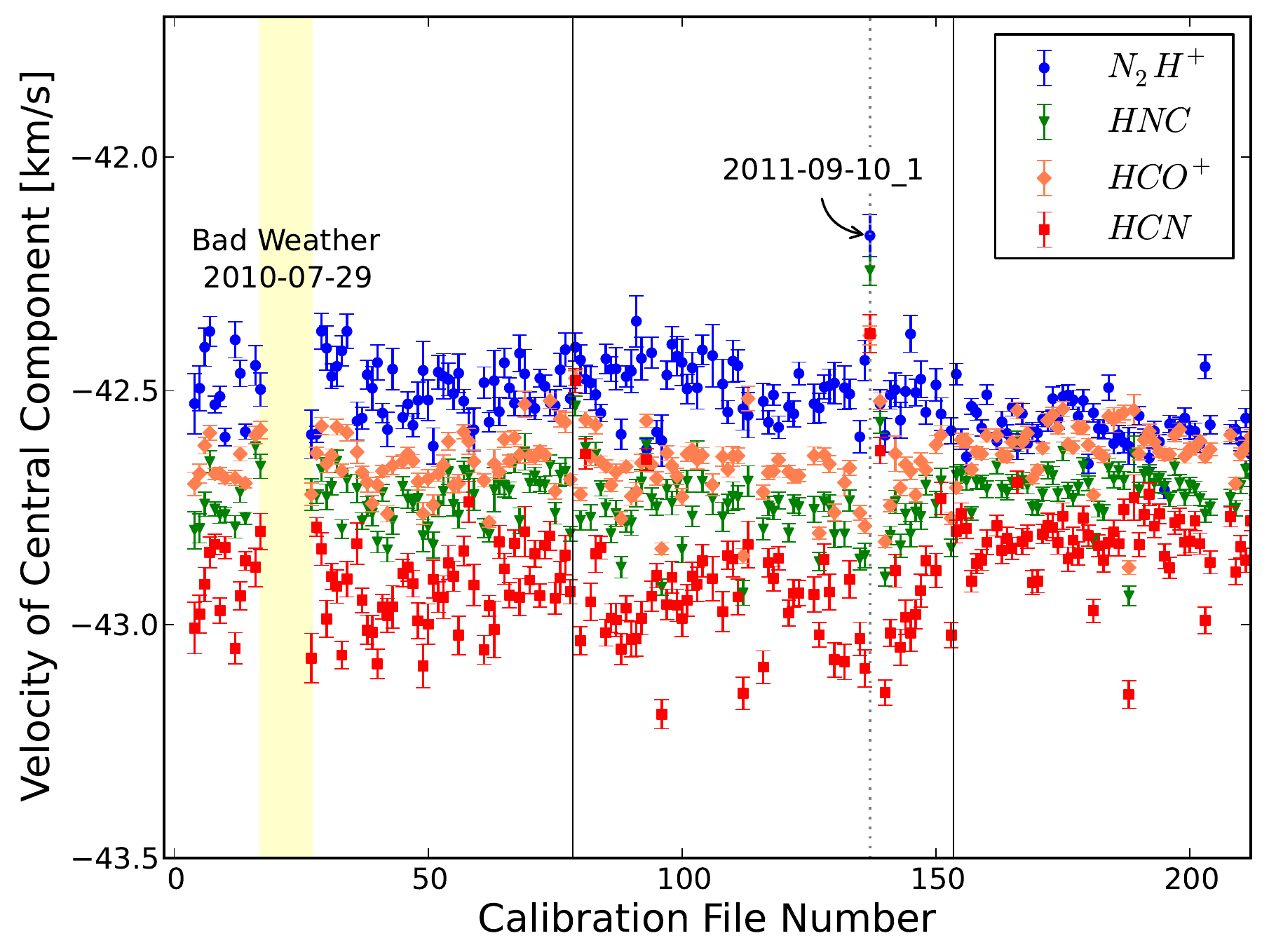}
\caption{The velocity of the central component for each of the four transitions as a function of sequential file number (effectively time). \nthp\ is shown as [blue] circles, \hnc\ is shown as [green] triangles, \hcop\ is shows as [orange] diamonds, and \hcn\ is shown as [red] squares. Black vertical lines delimit the breaks between our three  observing seasons. Additional features are marked and discussed in the text.}
\label{fig:time_series_vel}
\end{center}
\end{figure}

\begin{figure}[th]
\begin{center}
\includegraphics[width=\linewidth, angle=0]{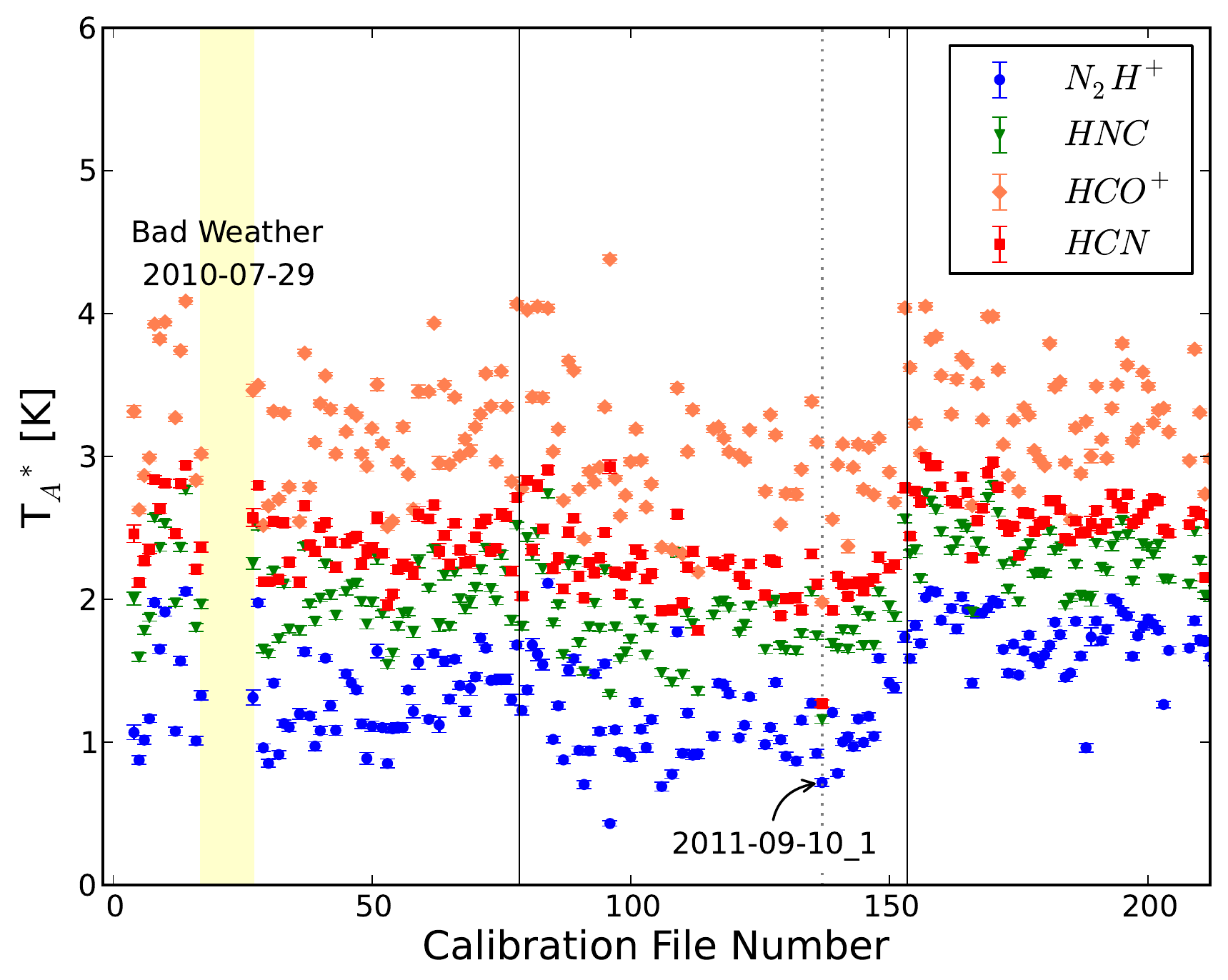}
\caption{The amplitude of the central component (if there are multiple components) for each of the four main transitions as in Figure~\ref{fig:time_series_vel}. }
\label{fig:time_series_int}
\end{center}
\end{figure}

Figures~\ref{fig:time_series_vel}~\&~\ref{fig:time_series_int} show the central velocity and the amplitude of the central component for each of the main four transitions. These figures only include days for which the fits to all four transitions met the reasonable fit criteria defined in Equations~\ref{eqn:reasonable-start}-\ref{eqn:reasonable-end}. These plots omit entries where the fit failed, based on criteria for reasonable parameters, but it does not specifically exclude data taken under poor weather conditions (as reported by high \Tsys\ values), although the criteria restricting the uncertainty on the fit parameters effectively eliminates data taken at high \Tsys.

Both figures exhibit a significant change near the start of the third season. This is not indicative of true source variability. Rather, during the third observing season, we started observations consistently at a later LST (local sidereal time), so that G301 was at a higher elevation and later time-of-day when it was observed. Variations in gain with elevation are common in radio telescopes and variations in gain as a function of time-of-day are also expected, particularly for a dish which is not temperature controlled (as is the case for the Mopra telescope dish); the gain variation is induced by variations in the thermal lag of structural members of the telescope \citep{Doyle:2009}. In general, the dish will tend to lose shape (and therefore efficiency) when the temperature has recently changed rapidly (shortly after dusk and dawn).

As discussed later, at low elevation there is evidence for a systematic offset in the pointing model, and this systematic offset accounts for the change in the behavior of the velocity in Figure~\ref{fig:time_series_vel}. This is a result of strong velocity gradients in our map of G301 which are significantly different for different transitions. 

The amplitudes and velocities of different transitions are highly correlated. That is, if the amplitude of \hnc\ is greater than average, the amplitude of \nthp\ will also be greater than average. We can quantify this with the correlation coefficient between the amplitudes of pairs of transitions. For example, the correlation coefficient between the amplitudes of \nthp\ and \hnc\ is 0.91, which is typical for the pairwise correlation coefficient of amplitudes in our data. This suggests that systematic trends in the gain of the Mopra telescope rather than purely random effects are producing the amplitude variation. 

We note two other features of these data. First, there is a significant gap between file number 17 and 27, which appears marked with a yellow band in Figures~\ref{fig:time_series_vel}~\&~\ref{fig:time_series_int}. These missing points correspond to a series of observations taken under poor weather conditions on 2010-07-29, with \Tsys $>$ 500 K. Normally we did not attempt to observe during such bad weather conditions but during this session, near the start of the survey, we continued to attempt to observe G301. Several of these observations resulted in detectable line emission, but the fits are often poorly constrained, and thus do not meet our quality criteria in Equations~\ref{eqn:reasonable-start}-\ref{eqn:reasonable-end} and are not included. 

The second feature of note is the behavior of points at file number 137, taken on 2011-09-10. These observations are labelled in Figures~\ref{fig:time_series_vel}~\&~\ref{fig:time_series_int} and a light gray dotted line is plotted to help guide the eye. This observation was taken during extremely windy conditions (wind speed 35 km hr$^{-1}$) and it is reasonable to think that there were larger than normal pointing error during these observations, resulting in the different velocity and decreased amplitude for the transitions (since the pointed position would be significantly off the peak of emission). This highlights the effect of high winds. 

The elevation range of the observations of G301 is strongly bimodal, clustered between 35\arcdeg\ and 40\arcdeg\ for observations taken at the start of most observing sessions during seasons one and two, and clustered between 55\arcdeg\ and 60\arcdeg\ for observations taken at the start of observing sessions which started later (including all of season 3). A couple of observations were made at very low elevation (25 - 30\arcdeg) when the shift started earlier than normal, and there are a few observations at intermediate elevations when variations in the schedule, bad weather, or mechanical failure caused us to start observing later. The data are also poorly sampled throughout all times-of-day. Most observing sessions started during the day due to the rise time of the Galactic plane during our observing season, Consequently, there are very few observations of G301 during the night. 

\begin{figure}[t]
\begin{center}
\includegraphics[width=\linewidth, angle=0]{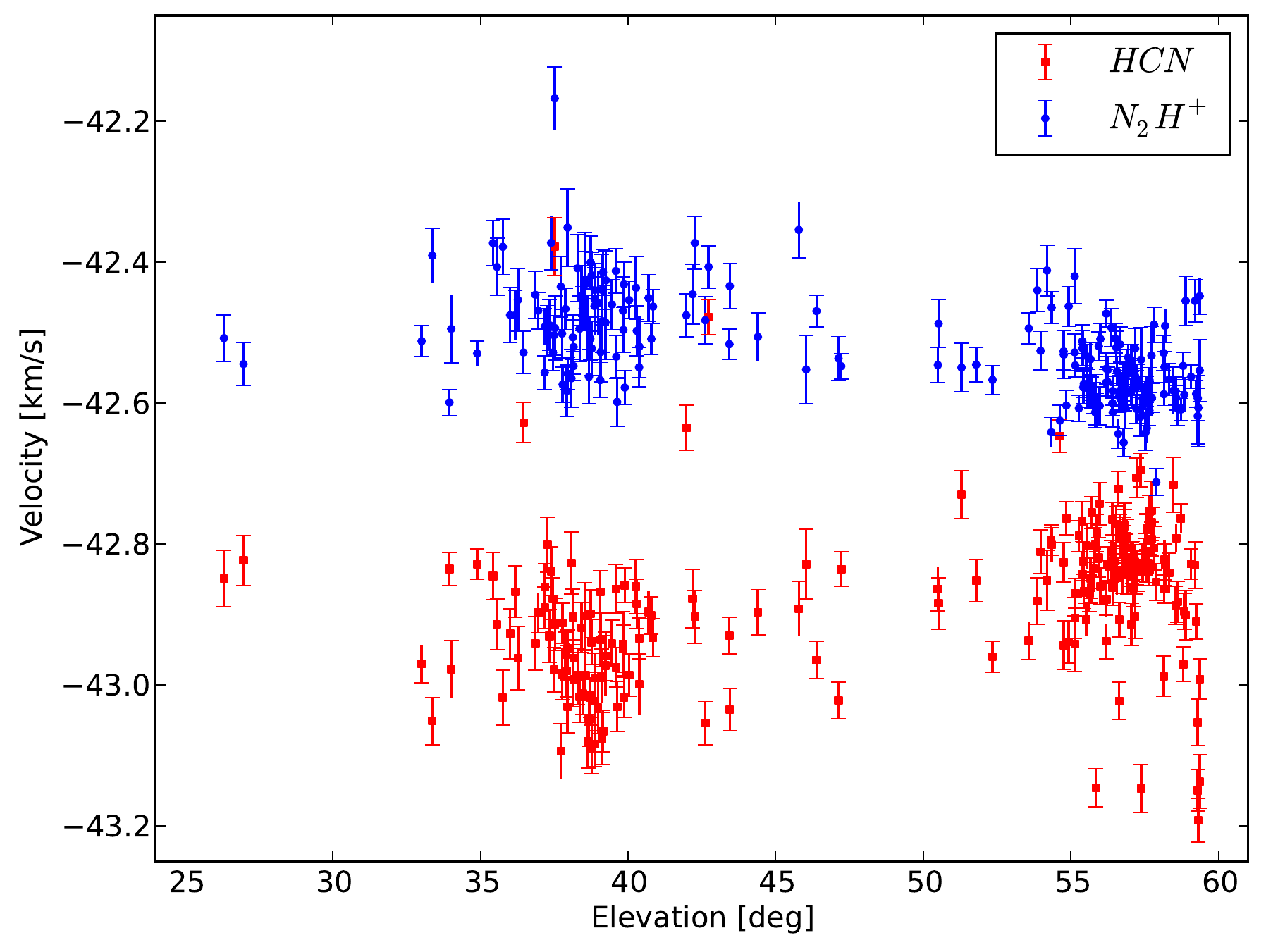}
\caption{The velocity of the central component for \nthp\ ([blue] circles) and \hcn\ ([red] squares) as a function of elevation. Significant trends are seen in both \nthp\ and \hcn.}
\label{fig:elevationvel}
\end{center}
\end{figure}

There is no systematic trend of peak parameters with \Tsys. Because G301 was typically observed at higher elevation during the third observing season, the trends seen in Figures~\ref{fig:time_series_vel} \& \ref{fig:time_series_int} can be partially ascribed to gain variation with elevation. Figure~\ref{fig:elevationvel} shows the velocity of \nthp\ and \hcn\ as a function of elevation. These transitions both show a change in velocity with elevation, but in the opposite sense.

It is not possible to disentangle the effects of varying gain as a function of elevation and pointing uncertainty without additional information. Both will tend to decrease the amplitude of the detected emissions, since pointing uncertainty will tend to scatter the observed position away from the brightest point in the source.

\subsection{Mapping Data}

We have three on-the-fly maps of G301. We cross-correlate integrated intensity images of these maps in order to determine their relative positional offsets. We use integrated intensity images of \nthp, \hnc, \hcop\ and \hcn\ and take the median offsets. The primary purpose of the cross-correlation is to allow us to optimally co-add the maps and produce a map with higher signal-to-noise for comparing with the PSW data. If the maps were simply co-added based on the positions recorded by the telescope, any errors in pointing would produce a smeared beam in the resultant map.

We shift the maps to align with the map taken on 2012-06-29. This map was taken at higher elevation, and thus we assume it will have the lowest absolute pointing uncertainty; we do not have an absolute position reference, but we do not need one for this analysis. Cross-correlation finds the following positional offsets: relative to the 2012-06-29 map, the 2011-05-06 map needs to be shifted by $-$2.7\arcsec\ in Galactic longitude and $+$2.25\arcsec\ in Galactic latitude; the 2011-08-22 map needs to be shifted by $+$1.8\arcsec\ in Galactic longitude and $-$3.6\arcsec\ in Galactic latitude.

\section{Analysis}
\label{sec:analysis}

Our analysis consists of two distinct steps. First, we use the velocities of the PSW observations to find the most likely location of each observation within our map of G301. This allows us to estimate the pointing reliability of our PSW observations. Second, for each PSW observation we examine the difference between the observed amplitudes of the transitions and the amplitudes of the transitions at the most likely locations in the map to model. We use the variation in this difference to model and remove the dominant sources of systematic gain variation and to estimate the residual absolute flux uncertainty.

\subsection{Most Likely Location}

\begin{figure}[t]
\begin{center}
\includegraphics[width=\linewidth, angle=0]{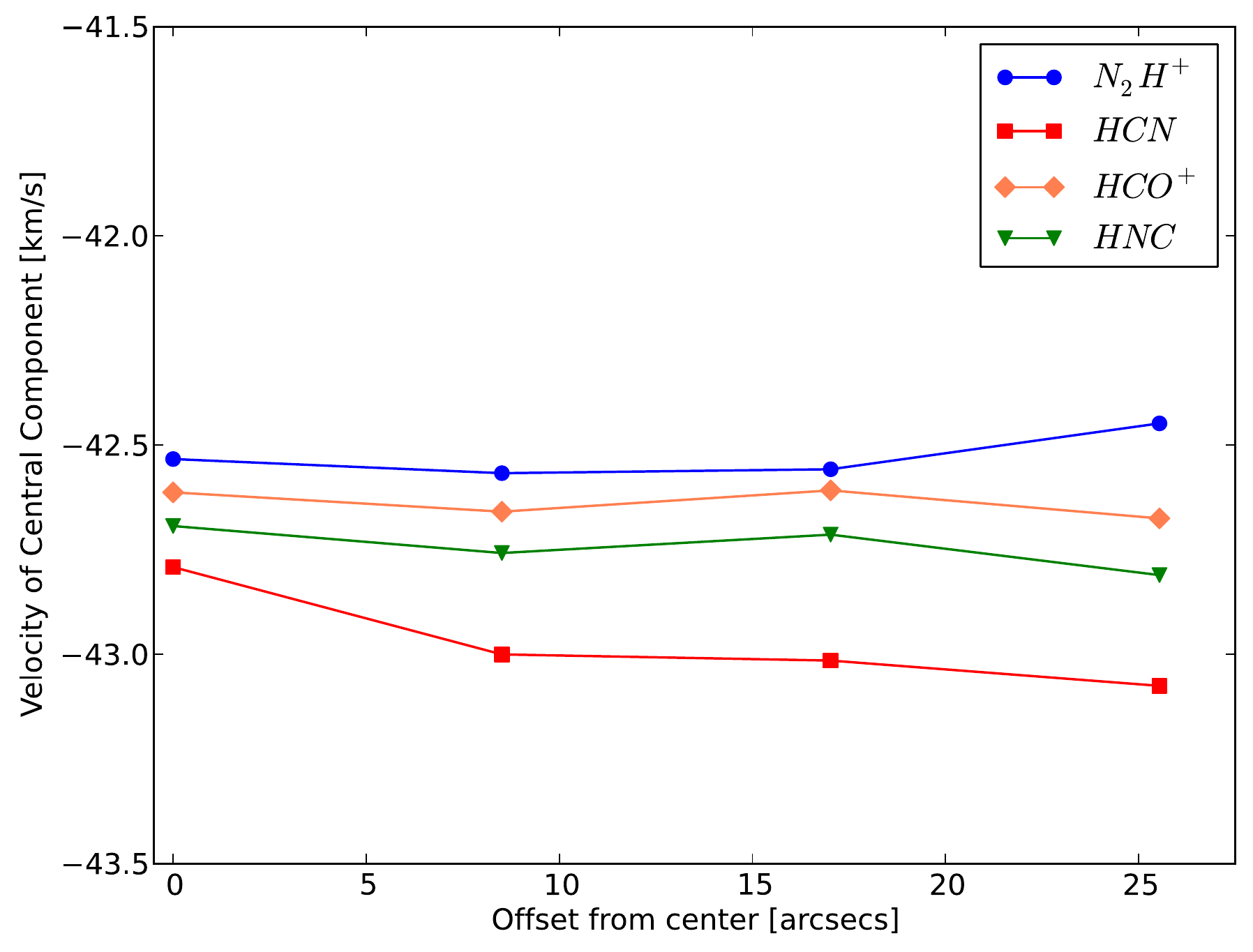}
\caption{The velocity of the central component for the four transitions along the direction shown in Figure~\ref{fig:bestpositions}, showing opposite gradients in \nthp\ and \hcn\ and relatively smaller changes in \hnc\ and \hcn.}
\label{fig:pv}
\end{center}
\end{figure}

\begin{figure*}[htbp]
\subfloat[][\nthp]{\includegraphics[width=0.48\linewidth]{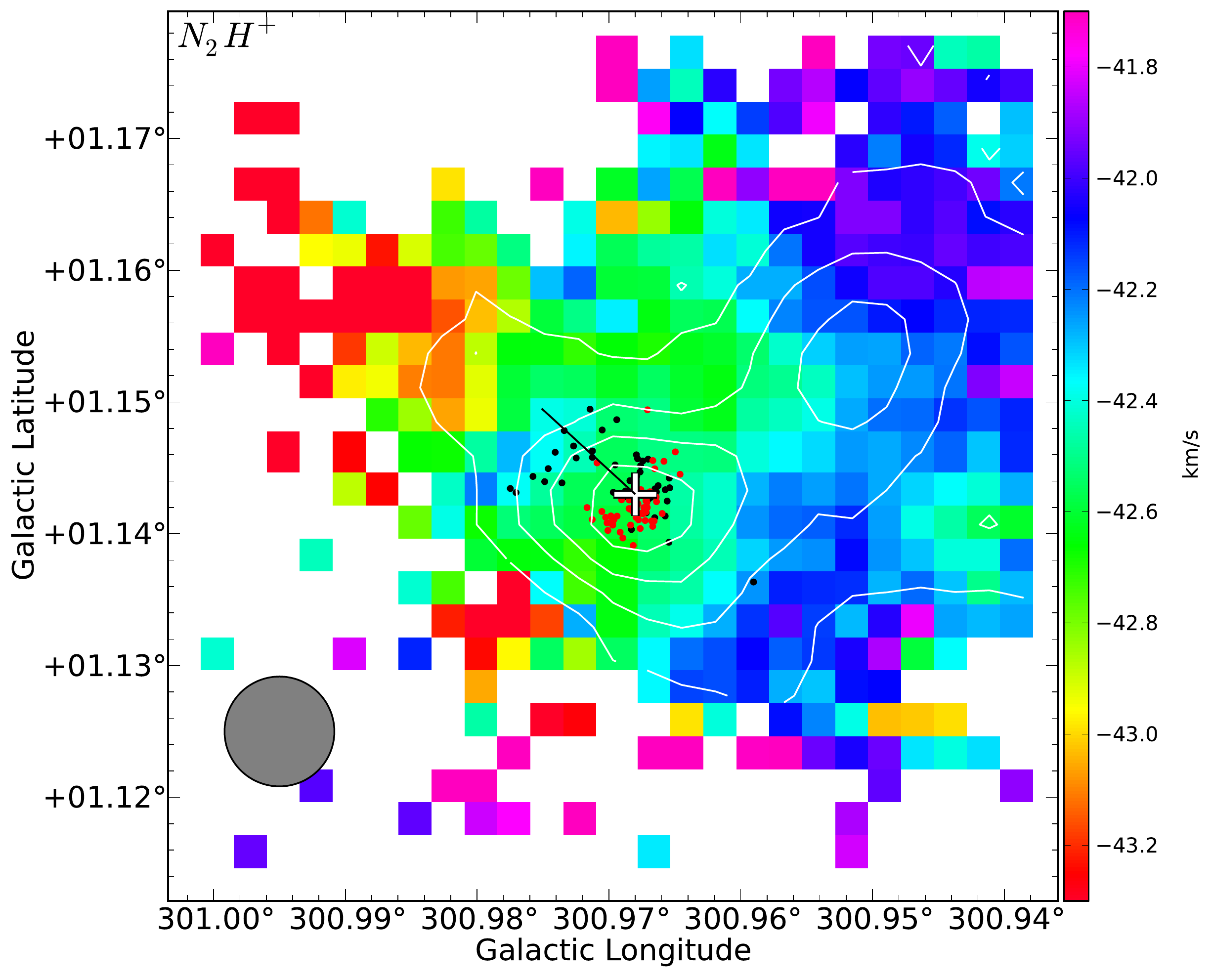}\label{fig:n2hp_positions}}
\qquad
\subfloat[][\hnc]{\includegraphics[width=0.48\linewidth]{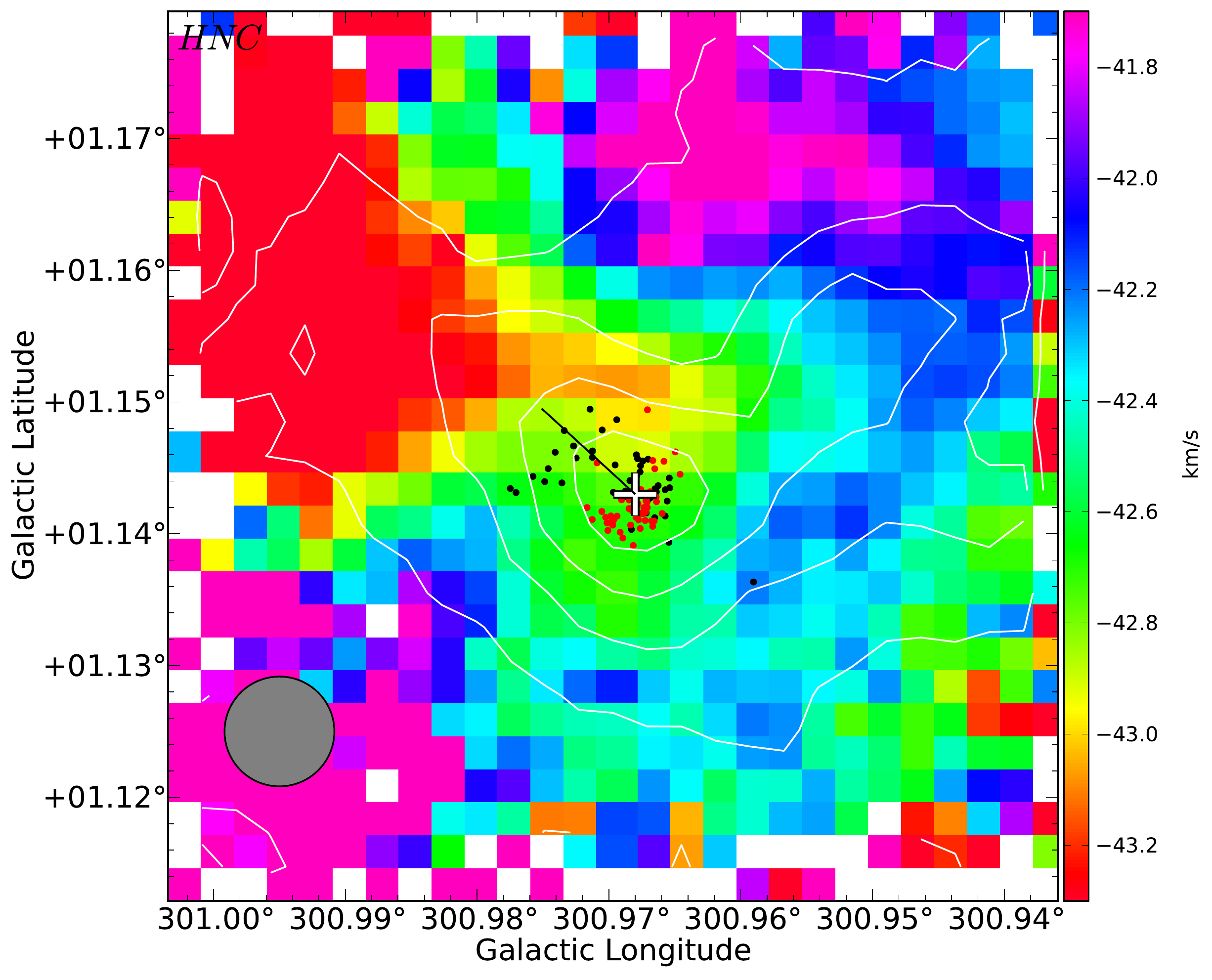}\label{fig:hnc_positions}}
\qquad
\subfloat[][\hcop]{\includegraphics[width=0.48\linewidth]{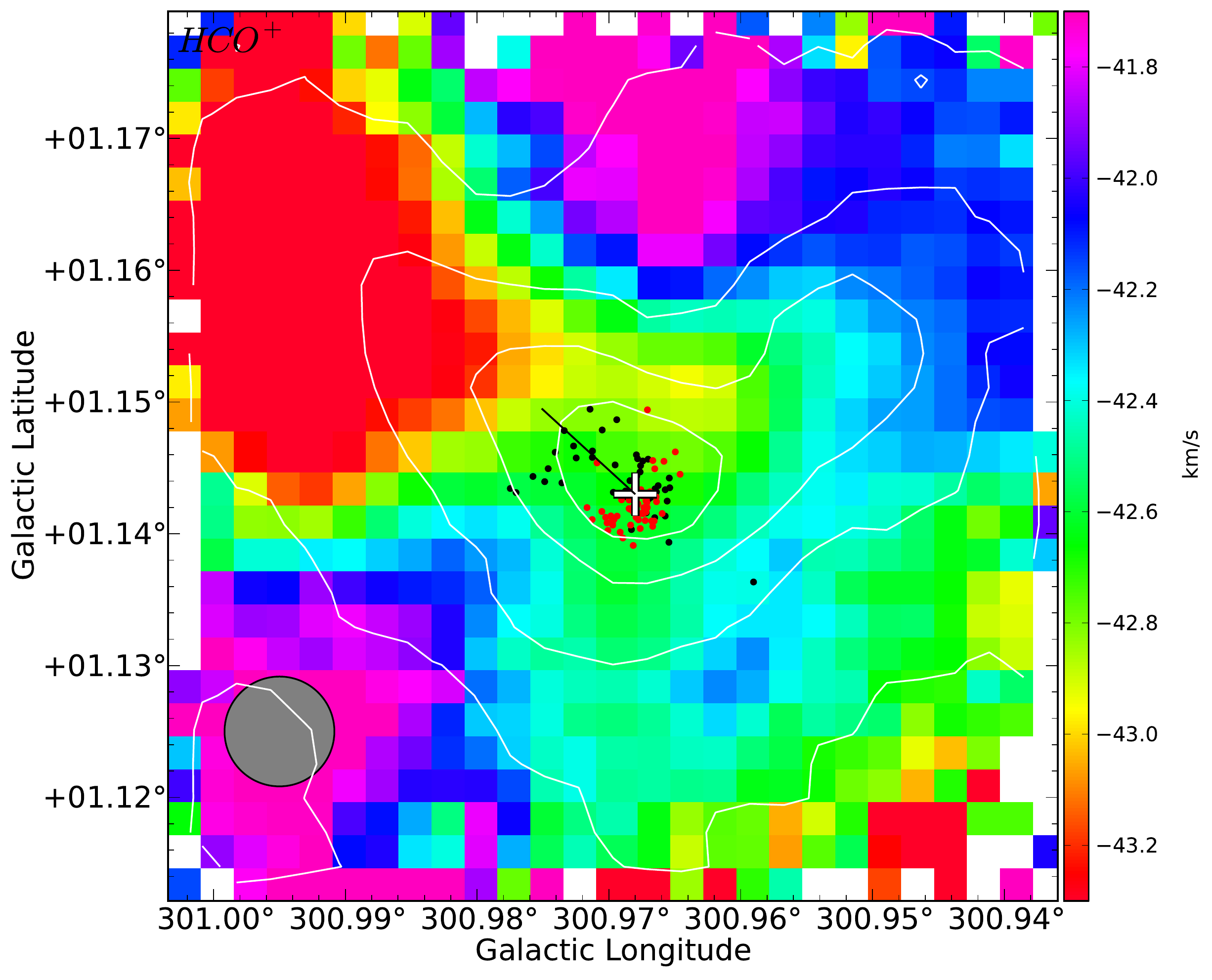}\label{fig:hcop_positions}}
\qquad
\subfloat[][\hcn]{\includegraphics[width=0.48\linewidth]{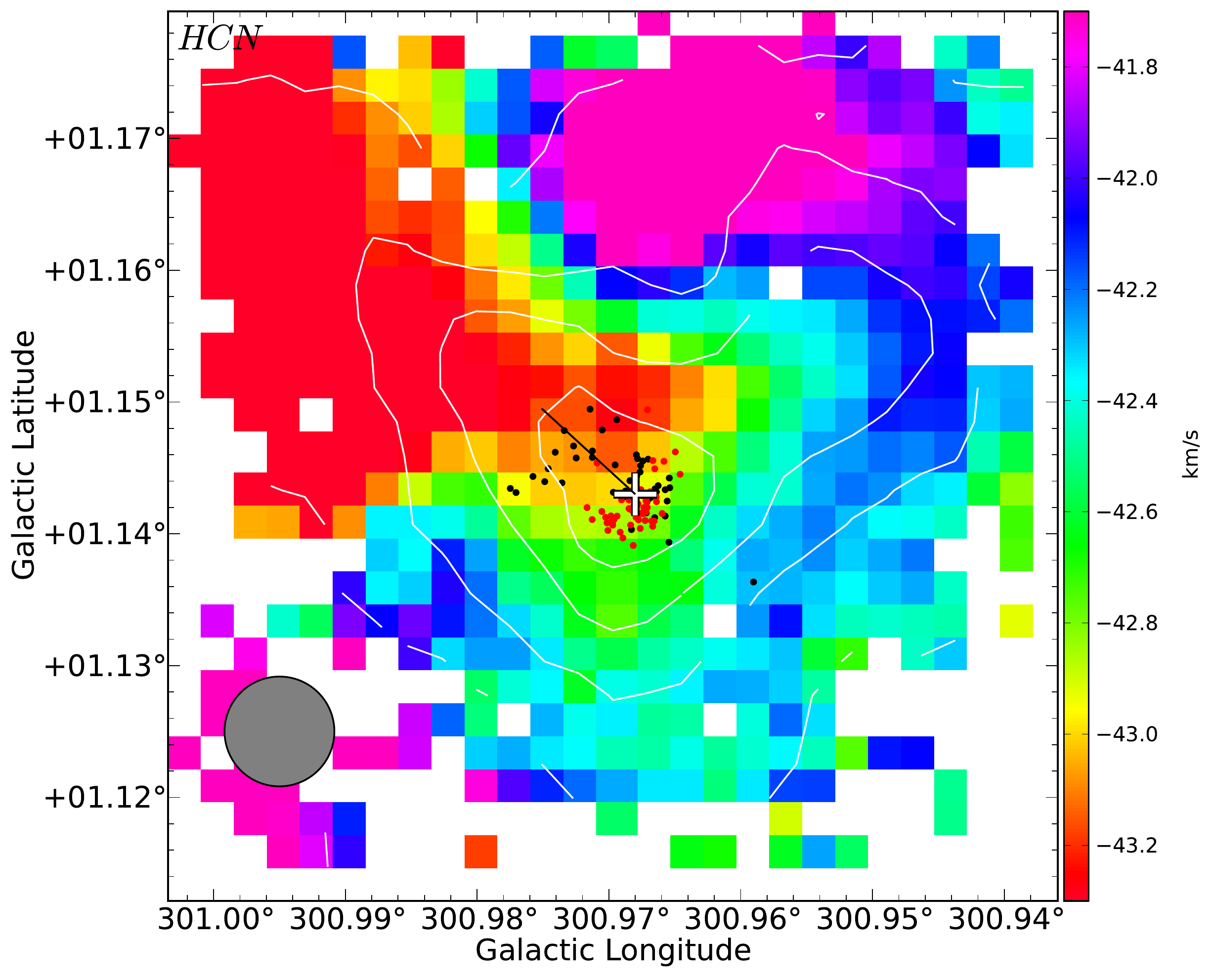}\label{fig:hcn_positions}}
\caption{The most likely positions of all PSW observations on a map of G301, derived by matching the velocities of all four main transitions simultaneously. In each subfigure, the color-scale shows the velocity of the central component of the transition, and the contours show the amplitude of the central component of that transition (with contours at 20, 40, 60 and 80\% of the maximum value). The black and red points show the most likely position of each observation, red for observations taken above 45\arcdeg\ of elevation and black for observations below that. A small amount of random jitter is added to each point to improve the display. The white cross shows the nominal position for the PSW observations. The beam of the Mopra telescope at 90 GHz is shown in gray. The black line shows the vector displayed in Figure~\ref{fig:pv}}
\label{fig:bestpositions}
\end{figure*}

We use the velocities of the four transitions in the PSW observations to find the most likely location of the observation within our co-added map of G301. That is, we use the velocities in isolation to estimate the pointing reliability. This is valid because systematic gain variations and absolute flux uncertainty will affect only the amplitude of the transitions, and not their velocity. In theory, intrinsic time variability of the source could cause changes in the velocity as well. In addition, errors in the reduction pipeline to derive velocity could contain a dependence on elevation; we examined the possibility that our reduction was incorrectly accounting for the Earth's rotation speed (the magnitude of this correction is elevation dependent) but found that this reduction was being performed correctly. Ultimately, the velocity offsets observed in Figure~\ref{fig:elevationvel} can be well understood by systematic pointing errors, and so we adopt this model as the simplest explanation. 

Qualitatively, there is a velocity gradient as one moves to larger Galactic latitudes and smaller Galactic longitudes away from the main clump. Figure~\ref{fig:pv} shows this gradient along the vector shown in Figure~\ref{fig:bestpositions}. This velocity gradient is large and in opposite directions for \nthp\ and \hcn, and relatively small for \hnc\ and \hcop. Therefore, an offset in our actual observed position between the points taken at low versus high elevation would produce the behavior seen in Figure~\ref{fig:time_series_vel}. 

To find the most likely location within the map of each PSW observation we seek the location within the map which minimizes the velocity offset of all four transitions simultaneously, subject to a reasonable pointing model. Specifically for our pointing model we assume that the average position of PSW observations coincides with the nominal targeted position. Furthermore, we assume that the pointing error is independent of angle ($\phi$) and therefore that the radial ($\rho$) distribution of PSW observations can be described by the Rayleigh distribution with a scale factor $\lambda$ so that
\begin{equation}
P(\rho) = \frac{\rho}{\lambda^2}e^{-\rho^2/2\lambda }.
\label{eqn:rho}
\end{equation}
The Rayleigh distribution describes the magnitude of a vector in two dimensions and is the two-dimensional equivalent of the Maxwell-Boltzmann distribution in three dimensions. 

Our Bayesian problem is therefore
\begin{equation}
P(\rho,\phi | data) \propto P(data | \rho,\phi) \times P(\rho,\phi),
\end{equation}
for each of $i$ data-points. We assume $\phi$ is uniformly distributed on the full range [0,2$\pi$] and that P($\rho$) is given by Eq.~\ref{eqn:rho} with $\lambda$ = 10\arcsec. This prior on $\rho$ comes from the May 15th, 2013 version of the Mopra Quick Reference Handbook\footnote{\url{http://www.narrabri.atnf.csiro.au/mopra/Mopra_QRH.pdf}} that estimates the global pointing model as having an 8.3\arcsec\ rms error in elevation and a 13.0\arcsec\ rms error in azimuth from an analysis of historical absolute pointing offsets. Ignoring this asymmetry for now, this corresponds to a $\lambda$ of 10.9\arcsec, which we round to 10\arcsec. Given $\phi$ and $\rho$ we assume that our data errors are Gaussian and well represented by the measurement error and thus 
\begin{equation}
P(data | \rho, \phi) \propto \sum\limits_{n} \left(\frac{v_{m,n}(\rho,\phi) - v_{p,n}}{\sqrt{\sigma^2_{p,n}+\sigma^2_{m,n}}}\right)^2,
\label{eqn:find-pos}
\end{equation}
for each ($n$) of the four transitions (\nthp, \hnc, \hcop, \hcn), $v_m$ refers to the velocity observed in a map pixel ($x$, $y$) at some distance $\rho$ and angle $\phi$ from the peak of the map, $v_p$ refers to the velocity obtained from the PSW observation, and $\sigma_p$ and $\sigma_m$ refer to the formal fit uncertainty on the velocity of the PSW observation and the velocity at a given map pixel respectively. 

We calculate Equations~\ref{eqn:find-pos} for each point in our maps of G301 and assign each PSW observation a most-likely position based on the map pixel that maximizes this probability. We do this both for the original maps and for interpolated maps where we interpolate down by a factor of four in both $x$ and $y$. We find that using the interpolated map allows us to well reproduce the observed PSW velocities.

\begin{figure*}[th]
\begin{center}
\includegraphics[width=\linewidth, angle=0]{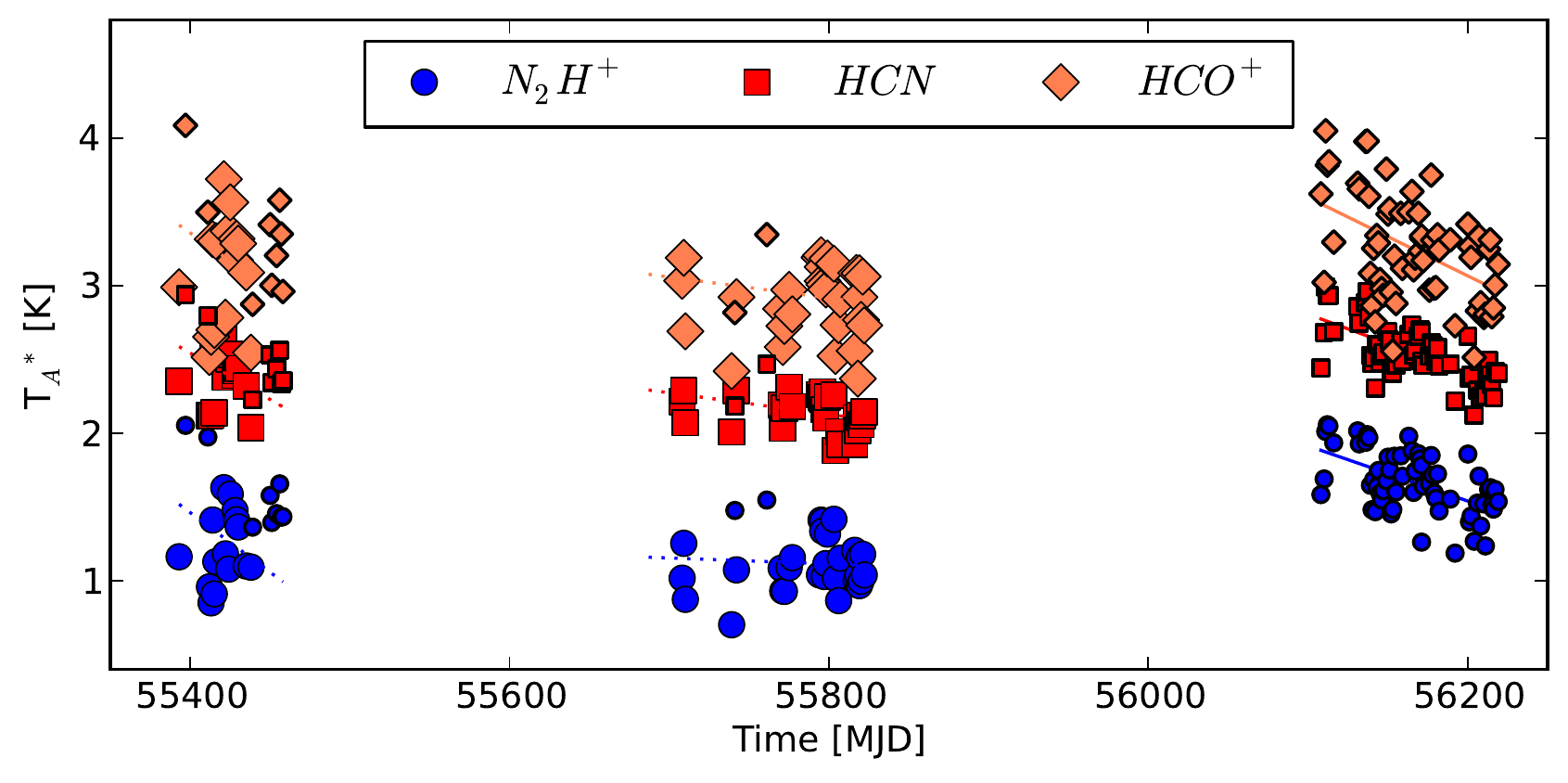}
\caption{The flux for three of our four main transitions (\hnc\ omitted for clarity) for PSW observations where \Tsys\ $<$ 180 K. The time range is displayed as the Modified Julian Date (MJD) and covers from July, 2010 to October, 2012. Points are coded by color and shape based on the transition observed, and are sized based on whether the observation was conducted at high elevation ($z > 45\arcdeg$; small points) or low elevation ($z < 45\arcdeg$; larger points). Lines show linear fits to the trend within each year. Large mean amplitude variations between observing seasons are apparent, as well as trends within each observing season. }
\label{fig:ampversustime}
\end{center}
\end{figure*}

The most likely location of each PSW observation is shown in Figure~\ref{fig:bestpositions} on each of the four main transitions. Some jitter is added to each point in this display in order to better visualize the density of points. These results show that the PSW observations taken at high elevation ($z > 45\arcdeg$) align quite well with the center of this map (effectively taken at high elevation, since maps were shifted to align with the map taken on 2012-06-29 at elevation 58-60 degrees), but that some of the points at lower elevation show a systematic offset toward larger Galactic latitude and longitude. The best-fit location for a small number of points are significantly further from the center of the map. These points correspond to PSW observations with significantly discrepant velocities; the PSW observation on 2011-09-10, highlighted in Figure~\ref{fig:time_series_vel} is one of these points.  

\subsection{Systematic Amplitude Variation}

With these most likely locations determined we proceed to consider the systematic amplitude variations seen in the PSW observations of G301. We assume that G301 has no significant intrinsic time variability and that all variation is due to gain variations. We expect that the Mopra telescope will experience some gain variation as a function of elevation, and also that the Mopra telescope may display some gain variation due to the fact that the dish is not kept at a constant temperature. This latter problem is exacerbated by the fact that observations were taken at different times during the day as well as over the course of many months during one observing season. The Mopra telescope could therefore be changing shape as the Sun warms the dish each day and as the ambient temperature changes during the season. One source of systematic variation can be discounted; no pointing model changes or receiver re-calibrations were performed during these three observing seasons. 

We perform this analysis in two parts. First we examine the data to find the dominant systematic variations. The goal is not to fully explain the absolute flux variation, but to identify the major systematic variations and quantify the remaining absolute flux uncertainty. Based on this examination we construct a hierarchical Bayesian model which allows us to coherently account for multiple sources of uncertainty and gain variations at the same time. 

\subsubsection{Examination of Amplitude Variation}

\begin{figure*}[th]
\begin{center}
\includegraphics[width=\linewidth, angle=0]{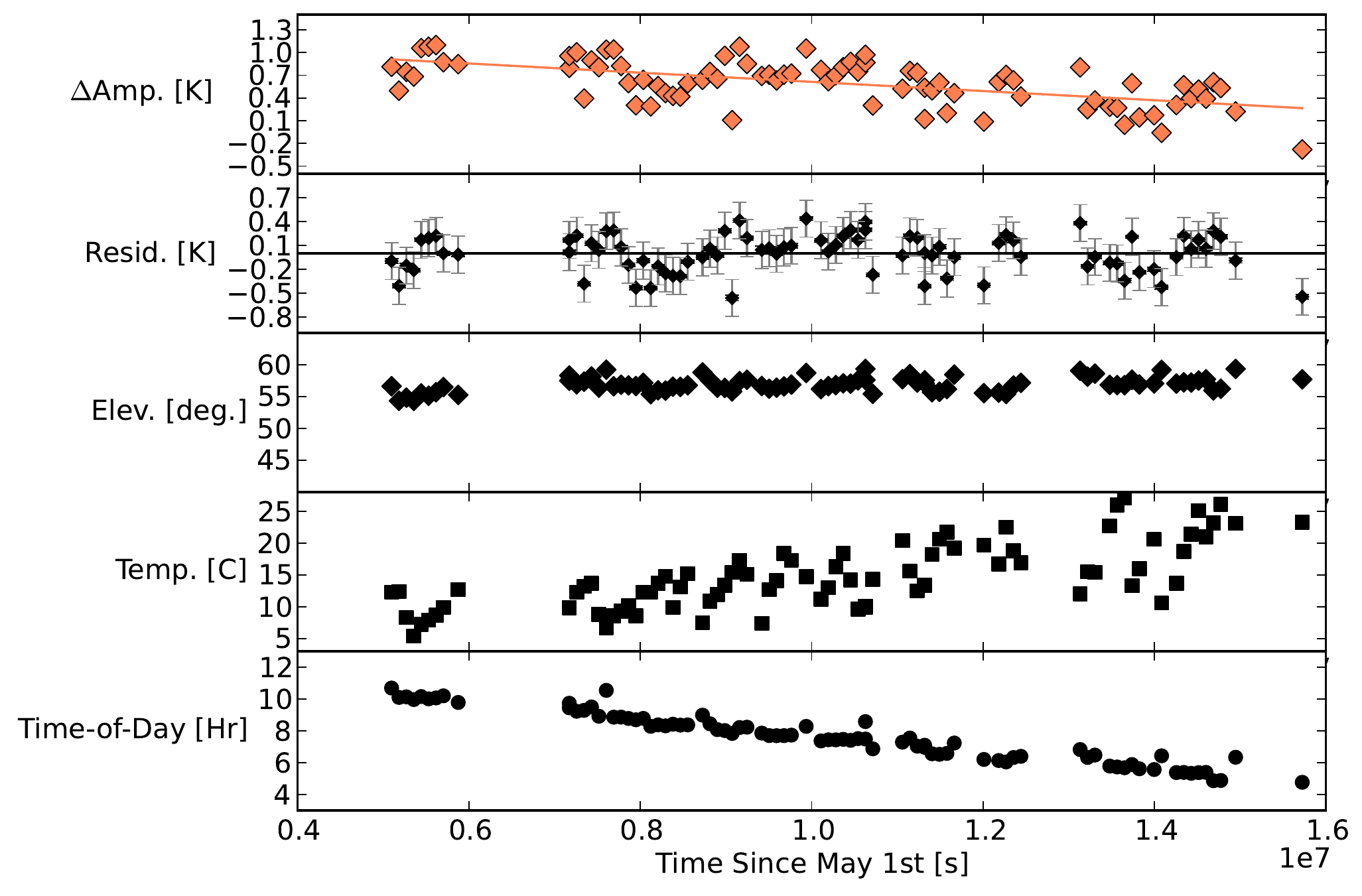}
\caption{A closer examination of the time variability of the \hcop\ transition amplitude during the third observing season. [Top Row] The difference between the amplitude of a molecular transition in a PSW spectrum and the amplitude of that transition most-likely location within the reference map. [Second Row] Residuals of the above linear fit; gray error bars show the inferred absolute flux uncertainty. [Third Row] Elevation is roughly constant. [Fourth and Fifth Rows] There is a trend with ambient temperature and with the time-of-day; the correlation is better with time-of-day. }
\label{fig:hcop_focus}
\end{center}
\end{figure*}

In order to simplify our search for the dominant systematic variations, we examine only the subset of our data taken during good observing conditions (\Tsys\ $<$ 180 K) and for which the best-fit position determined above is within 21\arcsec\ of the center of the map (which corresponds to where the amplitude of the transition is affected by less than 10\% due to pointing error). These data are shown in Figure~\ref{fig:ampversustime}. 

There are significant variations within each observing season as well as significant jumps between observing seasons. Within each observing season we fit a line to the amplitude of the molecular transition versus time. The amplitude decreases during the first and second season, but not with a high degree of statistical significance (1-2$\sigma$). In contrast, the decline is highly significant (6-$\sigma$) during the third season. All transitions show similar slopes. The fits for the first and second observing season are fit only to the points taken at low elevation, so as to avoid any variation produced by gain-elevation effects. The magnitude of the jump in amplitude from season two to season three is large (0.5 K), but the data taken during the third observing season were all taken at high elevation, in contrast to the previous two seasons. 

One explanation for these trends is that the gain of the Mopra telescope at 90 GHz is decreasing during the course of an observing season. During the first observing season this trend is partly obscured by the fact that later in the season we began observing sessions later, and thus observed G301 at higher elevation. A dependence of the gain on elevation is physically well-motivated and explains the rise at the end of the first observing season and the relatively higher transition amplitudes observed during the third season. 

The third observing season contains particularly robust evidence for a decrease in transition amplitude from the start to the end of the observing season. Figure~\ref{fig:hcop_focus} shows some possible explanatory variables, including elevation, temperature, and time-of-day (calculated as number of hours since sunrise). This figure shows difference between the PSW amplitude and the amplitude of the map at the most-likely location versus the time (in seconds) since May 1st. We choose May 1st as our reference time for a season so that all three observing seasons can be put on the same time axis when examining seasonal trends. This difference is offset from zero and linearly decreasing with time; error bars on the amplitude are smaller than the plot symbols. Linear fits to these relations show no significant structure in the residuals. This trend is not due to changes in elevation; our observing session start times were at roughly constant local sidereal time, so that G301 was at a similar elevation throughout the observing season. There is a negative correlation with ambient temperature ($\rho$ = -0.41) and a positive correlation with the time-of-day ($\rho$ = 0.52); the correlation is stronger with time-of-day. From this, we conclude that the dominant variations in the telescope gain can be modeled as due to elevation and time-of-day (as a proxy for thermal deformation). 

One complication of using time-of-day to explain the observed decrease during an observing season is that the real physical explanation of the decreased gain is likely deformation of the telescope due to differential thermal lag between components. This deformation will typically be most significant shortly after dusk and dawn when the temperature is typically changing most quickly. We examined the ambient (air) temperature at the Mopra telescope site in the hours preceding each PSW observation, but were unable to find a variable based on fitting these temperature profiles that produced a good correlation with the amplitude trend seen in Figure~\ref{fig:hcop_focus}. One possible reason for this is that the temperature of the dish (which is not measured directly) is strongly influenced by illumination by the Sun, and time-of-day is the best proxy available for this effect.

The third observing season data shown in Figure~\ref{fig:hcop_focus} do not include any points taken more than 12 hours after sunrise (after the trimming of low-quality data described in \S~\ref{sec:pointeddata}); the handful of points taken during the night are from the other two observing seasons. We thus do not have adequate coverage during the night to model this relationship, and therefore we exclude these points and focus on those taken less than 12 hours after dawn, where our data provide good coverage.

\subsubsection{Model}

For any given transition, we assume that the measured amplitude for any PSW observation, $a_{p}$ is generated as
\begin{equation}
a_{p} = \tilde{a}_{m} \times \eta(t) \times \zeta(z) + \epsilon_{p} + \epsilon_{f},
\end{equation}
where $\tilde{a}_{m}$ is the true amplitude of the transition at the position where the telescope was pointed within the map (as determined from minimizing Equation~\ref{eqn:find-pos}), $\eta(t)$ is the gain factor as a function of time-of-day ($t$), $\zeta(z)$ is the gain factor as a function of elevation ($z$),  $\epsilon_{f}$ is the remaining absolute flux uncertainty, and $\epsilon_{p}$ is the measurement error for the amplitude of an individual PSW observation. We assume that  $\epsilon_{f} \sim N(0,\sigma^2_{f}) $ and that $\epsilon_{p} \sim N(0, \sigma^2_{p}$) where $\sigma_{p}$ is our estimate of the uncertainty on $a_{p}$ and $\sigma_{f}$ characterizes the absolute flux uncertainty.

We further assume that both gain factors are linear functions of their dependent variables and normalized such that they are equal to unity at the elevation and time-of-day of our reference map ($t_{0}$, $z_{0}$) so that
\begin{equation}
\eta_{n}(t) = 1 + \beta_{n} \times (t_{p} - t_{0}),
\label{eqn:eta}
\end{equation}
and
\begin{equation}
\zeta_{n}(z) = 1 + \delta_{n} \times (z_{p} - z_{0}),
\label{eqn:zeta}
\end{equation}


where $t_{p}$ and $z_{p}$ are the time-of-day and elevation of each PSW observation. The elevation and time-of-day are the same for each of the $n$ species, but the model allows for different gain factors for each of our four main transitions, hence the subscripts on $\beta_n$, and $\delta_n$.

In the case of elevation, gain-elevation effects are often represented by a more complex function, since efficiency normally peaks around $z$ = 45\arcdeg-60\arcdeg, and drops at higher and lower elevation. However, the data are strongly clustered in two narrow elevation ranges, so a higher-order function can not reliably be fit. This fit should be used cautiously, and certainly not extrapolated to elevations outside of the measurements (i.e. $z$ $>$ 60\arcdeg\ or $z$ $<$ 30\arcdeg). Likewise, the data only cover between 3 and 12 hours after sunrise; over this period of time the gain of the Mopra telescope appears to respond roughly linearly, but this fit should not be extrapolated outside of this time range. 

Unfortunately, we do not have $\tilde{a}_{m}$, only an estimate, $a_{m}$, from a noisy map.
\begin{equation}
\tilde{a}_{m} = a_{m} + \epsilon_{m},
\end{equation}
where $\epsilon_{m} \sim N(0, \sigma^2_{m})$, therefore we have the following generative model for each PSW observation, $a_{p}$,
\begin{equation}
a_{p} = [a_{m} + \epsilon_{m}] \times \eta(t) \times \zeta(z) + \epsilon_{p} + \epsilon_{f}.
\end{equation}

We use the map of G301 taken on 2011-08-22 as our reference map, as it was taken at values of $t_{0}$ and $z_{0}$ near the median of our PSW observations. Specifically, $t_{0}$ = 5.5 hours and $z_{0}$ = 44 degrees. Recall that we used the combined map to find the most-likely location of each observation based on matching velocities. Those positions (appropriately shifted) are used to look up the amplitudes in this single map of G301, which has a well defined $t$ and $z$ associated with it (which the combined map does not). 

We now compute inferences on our parameters of interest, $\beta_n$, $\delta_n$, and $\sigma_{f}$, which represent the gain corrections for elevation and time-of-day and the absolute flux uncertainty. We assume uniform priors for $\beta_n$, $\delta_n$, and $\sigma_{f}$. We use {\sc pymc}\footnote{\url{http://pymc-devs.github.io/pymc/}} to compute the posterior probability distribution for each of these parameters, using adaptive Metropolis-Hastings sampling \citep{Haario:1998}. The traces converge well, and the posterior probability distributions are symmetric and single-valued, allowing us to specify the results simply as approximate gaussians.

\section{Results}
\label{sec:results}

\subsection{Pointing Uncertainty}

The best estimate of the pointing precision comes from matching the velocities derived from the spectra of the PSW observations against the velocities across the map. This gives a median position of $l$,$b$ = (300.9678\arcdeg, 1.1440\arcdeg) and a radial scatter of 9.8\arcsec\ at $z < 45 \arcdeg$ and a median position of $l$,$b$ = (300.9678\arcdeg, 1.1421\arcdeg) with a radial scatter of 4.3\arcsec\ at $z > 45\arcdeg$. The Galactic positions given above are not absolute, but are relative to the map of G301 taken on 2012-06-09 at $z = 59$\arcdeg. This corresponds to an offset of 6.8\arcsec\ in Galactic latitude between the median positions at the two elevation ranges, and suggests that there could be a systematic bias in the pointing of the Mopra telescope at different elevations. These results are modestly dependent on our choice of prior on $\rho$ in Eq.~\ref{eqn:rho}. In particular, decreasing $\lambda$ to less than 5\arcsec\ removes the offset between observations at low and high-elevation as all best-fit locations are now forced to be quite close to the nominal pointing center. Increasing $\lambda$ to 20\arcsec\ has a small effect on our estimate of the pointing precision, increasing our estimate of the pointing uncertainty by 2\arcsec.

The estimate for pointing uncertainty is therefore not fully encapsulated in a single number. At low elevation ($z < 45\arcdeg$) we infer a random scatter of 10\arcsec, but with a systematic offset of about 7\arcsec. At high elevation ($z > 45\arcdeg$) the Mopra telescope is relatively more precise, with a pointing uncertainty of 6\arcsec. For the full set of points (at all elevations), the radial scatter in inferred positions is 8\arcsec.

\begin{figure}[t]
\begin{center}
\includegraphics[width=\linewidth, angle=0]{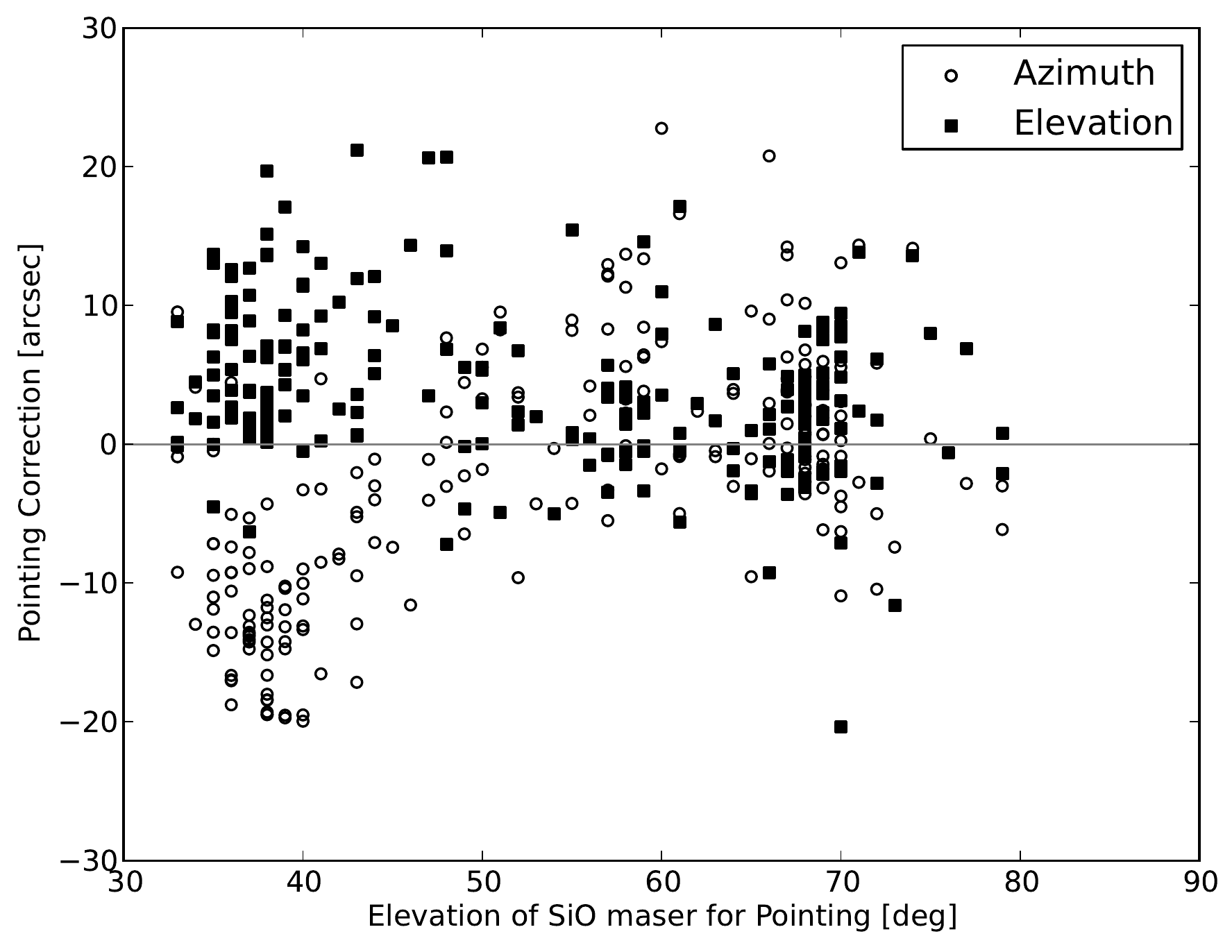}
\caption{Pointing corrections derived from observing an SiO maser immediately after observing G301. Corrections in both azimuth and elevation are significant below an elevation of 45\arcdeg, and small above this elevation.}
\label{fig:pointing-method-two}
\end{center}
\end{figure}

An additional check on these pointing results is made by considering the corrections to the pointing model required after a PSW observation of G301. As mentioned in \S~\ref{sec:PSWobs} we performed a pointing correction on an SiO maser immediately after a PSW observation of G301. The estimate of our pointing uncertainty derived in this fashion broadly agree with our results from matching velocities. Figure~\ref{fig:pointing-method-two} shows the corrections as a function of elevation of the SiO maser (all observations were in a fairly narrow range of azimuth). At low elevation ($z < 45\arcdeg$) there is a systematic correction of 6\arcsec\ in elevation and -11\arcsec\ in azimuth. The standard deviation of these corrections is 6\arcsec\ in elevation and 5\arcsec\ in azimuth. At higher elevation ($z > 45\arcdeg$) the average correction is small (2\arcsec\ in elevation and azimuth) and the standard deviation of these corrections is 7\arcsec\ in elevation and 6\arcsec\ in azimuth. The systematic pointing offsets at low elevation deduced from matching velocities correspond to offsets of 5\arcsec\ in elevation and -8\arcsec\ in azimuth (at the position of G301 and at the typical local sidereal time of G301 PSW observations), so this offset is in excellent agreement. The standard deviations of these corrections (expressed as radial corrections) are 8.2 and 8.8\arcsec\ for the two elevation ranges respectively. This is also in excellent agreement with our inferred overall pointing error of 9\arcsec. This analysis confirms our belief that the pointing is more accurate at high elevation.

Our estimate for the pointing uncertainty of the main MALT90 maps is 8\arcsec, comparable to the value of 10\arcsec\ often quoted for the Mopra telescope \citep[e.g.][]{Foster:2011, Jones:2012}. Since the pointing model seems to also be more accurate at high elevation, then it could be the case that MALT90 survey maps taken at high elevation have a pointing uncertainty of only about 6\arcsec.

\begin{figure*}[t]
\begin{center}
\includegraphics[width=\linewidth, angle=0]{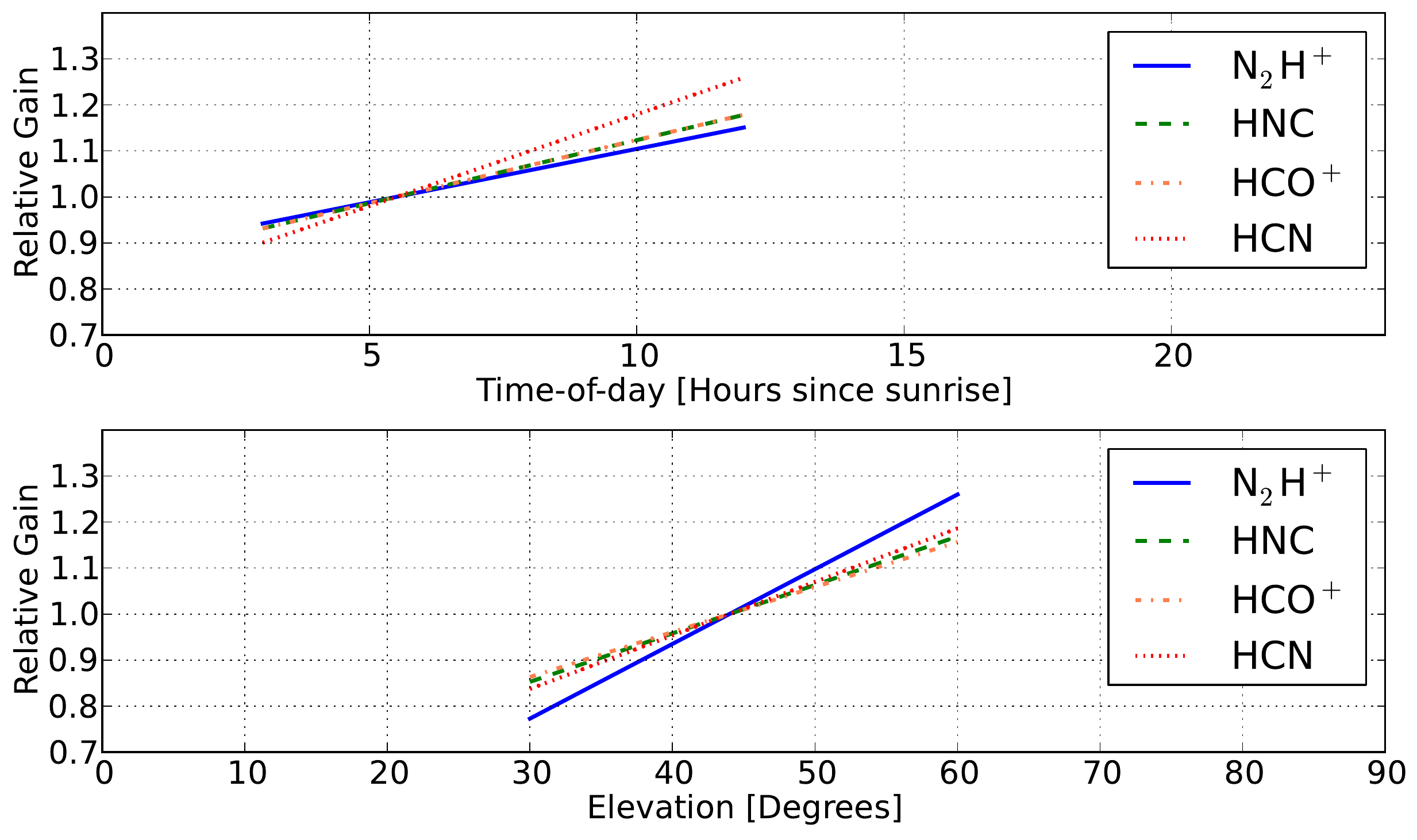}
\caption{Gain of the Mopra telescope at 90 GHz (normalized to unity at $t_0$ = 5.5 hours and $z_0$ = 44 degrees) as a function of time-of-day (top) and elevation (bottom) for the four main transitions in this study (\nthp, \hnc, \hcop, and \hcn). Relations are shown only for the ranges of parameters where they are calibrated by our observations of G301, and may deviate significantly from these linear fits outside of these ranges. }
\label{fig:gaincurves}
\end{center}
\end{figure*}

\subsection{Gain Factors}
\label{sec:gainfactors}
Table~\ref{table:gaincurves} lists our inferred parameters and 1-$\sigma$ uncertainties from our modeling of two factors influencing the gain of the Mopra telescope at 90 GHz, $\eta(t)$ where $t$ is the time-of-day (number of hours since sunrise) and $\zeta(z)$, where $z$ is the elevation in degrees. These relations are described in equations~\ref{eqn:eta} and \ref{eqn:zeta}. Both these gain relations are normalized to be one at the time-of-day and elevation of our reference map ($t_{0}$ = 5.5 hours, $z_{0}$ = 44 degrees), and describe how the gain of the Mopra telescope changes in our PSW observations taken at different times-of-day and elevations. Figure~\ref{fig:gaincurves} visualizes these relations over the ranges where the explanatory variables (elevation and time-of-day) are well sampled in our observations of G301.

\input{gainparameters_table.tex}

All the gain versus elevation relations ($\delta$ in Table~\ref{table:gaincurves}) are consistent with each other at the 3-$\sigma$ level. Taking \nthp\ as an example, the relation implies that at 30 degrees of elevation, the observed flux would be only 78\% of the baseline flux observed at 44 degrees, while the flux would be 126\% of the baseline at 60 degrees. The relation is not calibrated outside of this elevation range, and should not be used at lower or higher elevations. In particular, we expect that the gain might peak around 60 degrees of elevation (based on other telescopes) and thus extrapolating this linear relation to higher elevations would produce dramatically incorrect answers.

The gain variation with time-of-day relations are also (2-$\sigma$) consistent with each other. For \nthp, the slope of this gain relation is less than 3-$\sigma$ different from zero, but the slope is more than 4-$\sigma$ significant for all the other transitions. The effect is strongest in \hcn. This is somewhat counter-intuitive in our picture where temperature deformation of the dish changes the efficiency of the telescope. In particular, since \hcn\ is at the lowest frequency of our transitions, it would generally be the least sensitive to gain variation due to deformation of the dish (although the frequency difference is not large). Nonetheless, the relatively large uncertainties on these relations mean that the four relations are all consistent with each other. Relative to a baseline observation taken at 5.5 hours past sunrise, the model implies that for \hcop one would observe a flux of 118\% of the baseline flux at 12 hours past sunrise and a flux of 93\% of the baseline flux at 3 hours past sunrise. This relation is not calibrated outside of this range. 

\begin{figure*}[t]
\begin{center}
\includegraphics[width=\linewidth, angle=0]{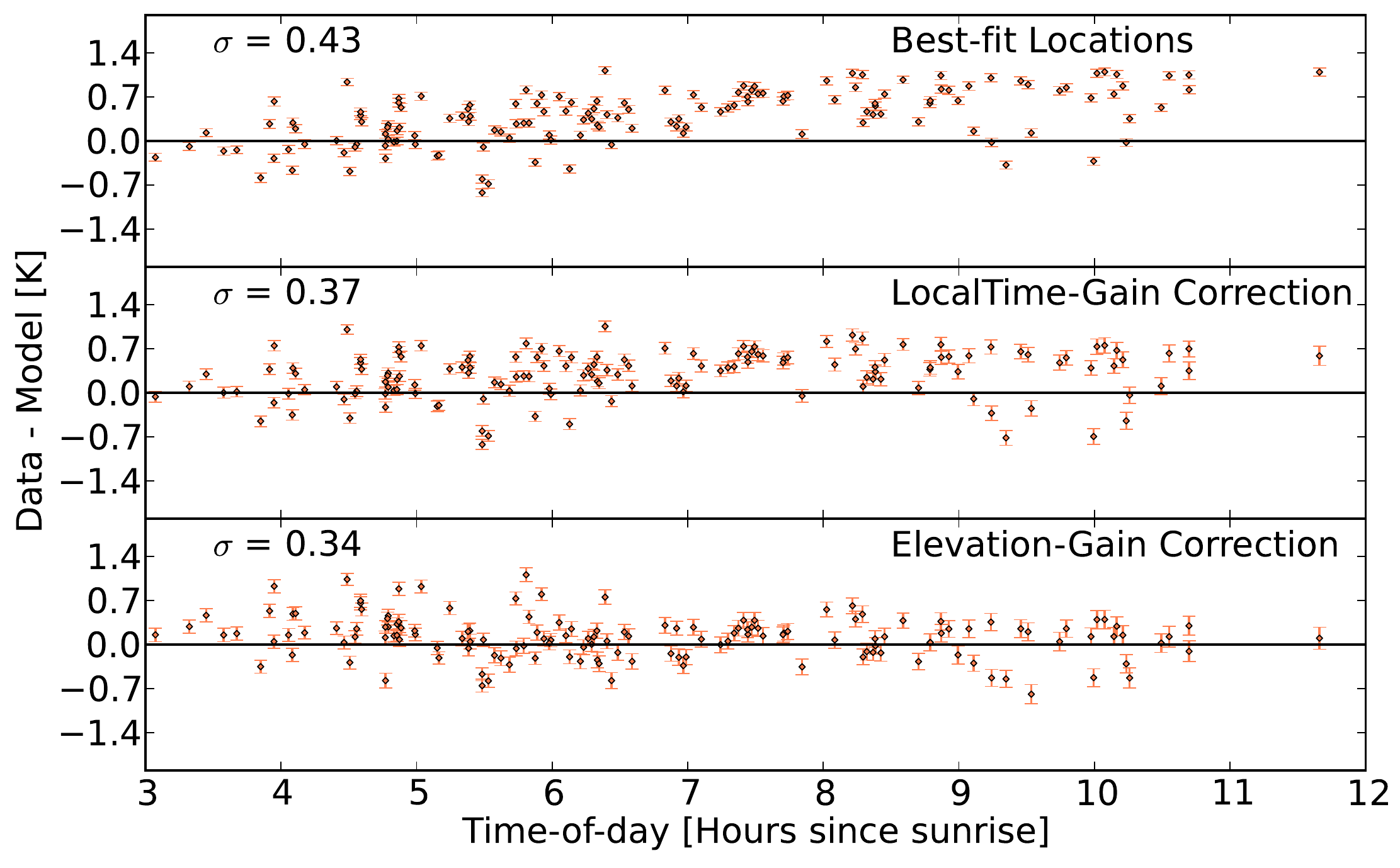}
\caption{Residuals of our model explaining the amplitudes of PSW observations of \hcop\ after including various refinements. [Top Row] The amplitude difference from the most-likely location in the map. [Middle Row] Residuals after including the gain variation with time-of-day. [Bottom Row] Residuals after also including the gain variation with elevation. }
\label{fig:model_check}
\end{center}
\end{figure*}

After accounting for these first-order effects, there is still more variation in the PSW observation amplitudes than can be accounted for by noise in the spectrum and its resultant uncertainty on the fitted amplitudes. Our model captures this number as an additive Gaussian noise term characterized by $\sigma_f$ in Table~\ref{table:gaincurves}, although this could also be modeled as a (multiplicative) variation in gain. The absolute magnitude of this variation is 0.24$\pm$0.1~K for \hnc\ and 0.26 - 0.36$\pm$ 0.2~K for \nthp, \hcn, and \hcop. Using typical transition amplitudes of \nthp = 1.5~K, \hnc = 2~K, \hcop = 3.5~K and \hcn = 2.5~K, these variations can be expressed as percentage variations of 17\%, 12\%, 10\% and 11\% respectively. This roughly follows our expectation that the absolute flux uncertainty would be a function of frequency, and be worst at high frequencies (i.e. the flux accuracy for \nthp at 93 GHz is worse than the other transitions between 88 and 91 GHz). 

By contrast, the following numbers describe the standard deviation of the PSW observation amplitudes without accounting for these first-order effects ($\sigma_i$ in Table~\ref{table:gaincurves}). For \nthp, $\sigma_{i}$ = 0.38~K, for \hnc, $\sigma_{i}$ = 0.34~K, for \hcop, $\sigma_{i}$ = 0.43~K, and for \hcn, $\sigma_{i}$ = 0.29~K. As percentage variations these are roughly 25\%, 17\%, 12\%, and 12\% respectively. Correcting for the first order changes in the Mopra telescope's gain at 90 GHz therefore produces a modest, but significant, improvement in the absolute flux calibration of the data. 

These flux uncertainties are much larger than the fitting uncertainty for our amplitudes in our PSW data (typical $\sigma_{a}$ = 0.03 - 0.05~K) and in our maps ($\sigma \sim$ 0.05 - 0.08~K for for a typical MALT90 map). This uncertainty sets a limit on the precision of our transition parameter determinations using MALT90 data. We expect that the absolute flux uncertainty is a slowly varying function of time. Therefore within a map, one does not have to take into account this absolute flux uncertainty when measuring relative quantities, such as the 50\% contour of emission in a given molecular transition within a source. In addition, because of the strong correlation among molecular transition amplitudes, we expect that this residual flux uncertainty is frequency independent (at least in sign, if not exactly in amplitude) and thus the fact that MALT90 maps of different species are made concurrently should remove most of this absolute flux uncertainty when looking at, for instance, molecular transition ratios from one source to another. 

Our model is shown in Figure~\ref{fig:model_check} which displays the molecular transition amplitudes from the PSW data minus the model amplitudes after including a series of refinements, which include: (1) taking the molecular transition amplitude from the most-likely location in the reference map, (2) including gain variation with time-of-day, and (3) including gain variation with elevation. This plot shows all three seasons of \hcop\ together, and the error bars include the uncertainty from our model parameters. The unweighted standard deviation of these points decreases with each refinement to the model.

\subsection{Intrinsic Source Variability}

\begin{figure}[t]
\begin{center}
\includegraphics[width=\linewidth, angle=0]{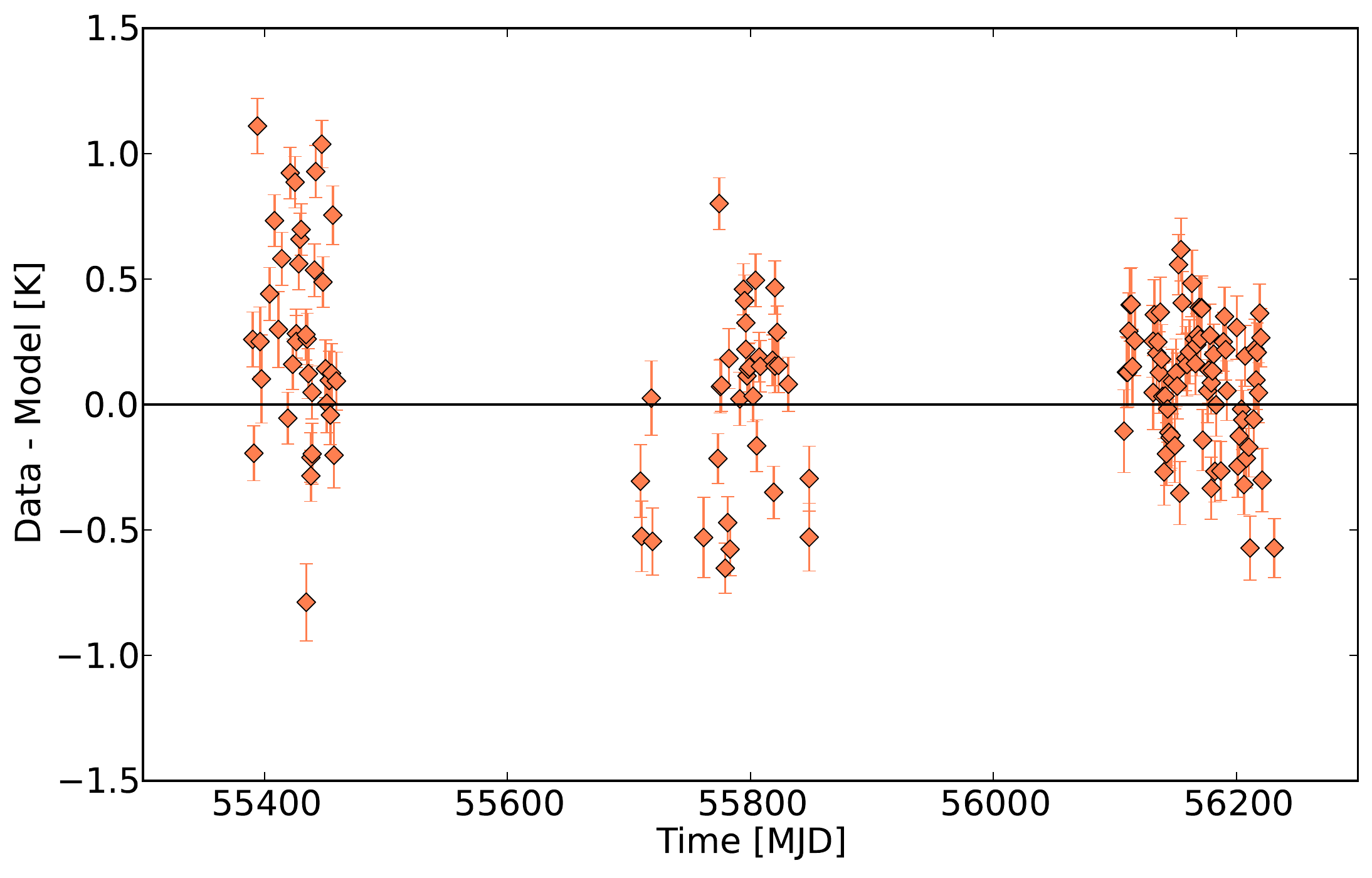}
\caption{Residuals of our model explaining the amplitudes of PSW observations of \hcop\ versus time expressed as MJD. This residual is equivalent to the bottom panel of Figure~\ref{fig:model_check}. Any intrinsic source variability is less than the magnitude of our systematic corrections. }
\label{fig:intrinsic}
\end{center}
\end{figure}

Figure~\ref{fig:intrinsic} shows that there is still some residual variation between seasons, although it has been much reduced (c.f. Figure~\ref{fig:ampversustime}). These variations would be statistically significant if not for the systematic nature of our gain corrections (\S~\ref{sec:gainfactors}). That is, comparing the mean and standard error on the mean for the second observing season (-0.16$\pm$ 0.02 K) and the third observing season (0.04$\pm$0.01 K) appears to show a statistically significant difference. However, since most of the third observing season spectra were taken at high elevation (55 - 60\arcdeg) while most of the second observing season points were taken at low elevation (35 - 45\arcdeg), the systematic correction for the elevation-gain relation is roughly 0.5 K, far larger than the residual difference. 

This variation could still hint at intrinsic source variability, but it could also be simply another instrumental systematic not fully modeled in this work. The underlying physical explanation for this variability could be systematically different in different observing seasons, and therefore intrinsic source variability at this level (0.3~K, or 10\%) cannot confidently be measured. The data would be sensitive to much larger intrinsic source variability, such as the 40\% (continuum) flux variation seen in an \UCHII\ by \citet[e.g][]{Franco-Hernandez:2004}. If G301 exhibited a similar flux variability in line emission, this would produce roughly a 1.4~K change in the brightness of the \hcop\ transition, a variation to which the data would be sensitive. Continued monitoring of this source will help us to constrain the magnitude of any intrinsic source variability.

\section{Conclusions}

We have presented the MALT90 data for this survey's characterization source, G301, a well-studied \UCHII. Repeated PSW observations of this source (at the start of each observing session over three seasons of the survey), in combination with a high-quality map of this source, allow us to characterize the system performance of the Mopra telescope at 90 GHz, and thus several parameters describing the repeatability of measurements and the pointing reliability of the MALT90 survey.

We see strong systematic time variability in the amplitudes of transitions in our single-point observations of G301, but we do not believe that this is due to intrinsic source variability. Instead, the long-term amplitude trends can be explained by a model in which the Mopra telescope at 90 GHz has two significant gain variations, one as a function of time-of-day (probably related to temperature fluctuations), and one as a function of elevation. To first order, the variation within an observing season is due to changes in the time-of-day as our observing session starts when the Galactic plane rises. The variation between the first two seasons and the third season is due to elevation (since the third observing season started at later local sidereal time). 

Our main results characterizing the survey are:
\begin{itemize}
\item We estimate our pointing uncertainty to be 8\arcsec. This number includes a systematic offset between observations taken at different elevations, with observations at low elevation likely to be mis-pointed. The pointing uncertainty is only 6\arcsec\ for sources observed above 45\arcdeg\ of elevation (the majority of MALT90 sources).
\item We quantify the gain-elevation relation for the Mopra telescope at 90 GHz (Table~\ref{table:gaincurves} and Figure~\ref{fig:gaincurves}). The strong clustering of our observations in two small elevation ranges (around 35-40\arcdeg\ and around 55-60\arcdeg) prohibits us from fitting anything of higher order than a linear relationship and limits the range over which such a correction can be applied.
\item We infer that the Mopra telescope at 90 GHz experiences gain variation as a function of time-of-day. In particular, efficiency increases linearly during the day from 3  to 12 hours after sunrise. This variation is not characterized outside of this time period. The gain relations are consistent for the four different transitions used in this analysis. 
\item After removing these two sources of gain variation, there is a remaining absolute flux uncertainty of 0.24 - 0.36 K, or 10 - 17\% depending on the transition in question. Without this correction, the absolute flux uncertainty is 0.29 - 0.43 K, or 12 - 25\%. This systematic uncertainty dominates over the noise inferred from examining signal-free sections of the spectra. For certain applications, this sets the uncertainty of MALT90 molecular transition amplitudes, although for others (i.e. line ratios) the strong correlations among transition amplitudes and the fact that all the molecular transitions are observed simultaneously minimizing this source of uncertainty.
\end{itemize}

We do not use our estimates of these systematic gain variations to correct the fluxes in the MALT90 survey. The primary reason is that our observations of G301 do not adequately span the elevation and time-of-day ranges present in our full dataset; a correction of the full set of survey maps would therefore involve significant extrapolation. In particular, our observations of G301 only well sample two small ranges in elevation (35-40\arcdeg and 55-60\arcdeg) and only well sample the range from a few hours after sunrise to just after sunset. 

A second reason is that not all MALT90 maps can be characterized by a single elevation or time-of-day. Although most sources were observed during a contiguous block of time (that is, the map scanning in Galactic latitude immediately followed the map scanning in Galactic longitude), for some sources we observed the two different scan-maps at discontiguous times for a variety of reasons. The most common cause was only finishing a scan-map in one direction for the last source of a given observing session. Under the normal data reduction pipeline, maps are combined with \Tsys\ weighting; to apply the elevation and time-of-day corrections presented here would require an additional weighting factor before co-addition. For both these reasons, we present the MALT90 data without these corrections applied. Nevertheless, these corrections are important to understand for reliable analysis and interpretation of MALT90 data and we encourage their use where appropriate. 

The characterization of telescope parameters such as the pointing uncertainty, absolute flux calibration, and gain-variation relations will be useful for other users of the Mopra telescope at 90 GHz, since MALT90 observes in a fairly standard fashion. We will continue to monitor G301 as part of the MALT90 survey, including observations at a broad range of elevations and times-of-day, and present updated values for these parameters with the final data release paper. In addition, increased observations will hopefully allow us to break the degeneracies among observing season, elevation, and time-of-day and thus place strong upper limits on any intrinsic variability of molecular transitions in \UCHII s such as G301.

\section*{Acknowledgments} We thank the referee, Michael Burton, for useful suggestions which improved the paper. The Mopra Telescope is part of the Australia Telescope National Facility and is funded by the Commonwealth of Australia for operation as a National Facility managed by CSIRO. The University of New South Wales Mopra Spectrometer Digital Filter Bank used for the observations with the Mopra Telescope was provided with support from the Australian Research Council, together with the University of New South Wales, University of Sydney, Monash University and the CSIRO. The authors would like to thank the staff of the Paul Wild Observatory for their assistance during these observations. The MALT90 project team gratefully acknowledges the use of dense clump positions supplied by ATLASGAL. ATLASGAL is a collaboration between the Max Planck Gesellschaft (MPG: Max Planck Institute for Radioastronomy, Bonn and the Max Planck Institute for Astronomy, Heidelberg), the European Southern Observatory (ESO) and the University of Chile. J.M.J gratefully acknowledges funding support from NSF grant AST-1211844.

\appendix

\section*{Appendix: PSW Observations}

Table~\ref{table:source_info_long} presents a summary of every PSW observation of G301, including the file number, the name of the file (which incorporates the UT date when the observation was started) and the parameters important in assessing pointing reliability and gain variations. These include the azimuth, elevation, the time (listed as modified Julian date or MJD), the time-of-day (hours since sunrise), the ambient temperature at the time of observations, and the time since May 1st within each observing season (our proxy for time-of-year). 

Table~\ref{table:fit_parameters_long} shows the fit parameters (velocity and amplitude) with uncertainty for the central components of each of the four main transitions. Parameters are only shown if the fit was reasonable according to the criteria given in Eq.~\ref{eqn:reasonable-start}-\ref{eqn:reasonable-end}. Our analysis excludes the data taken on the first day of the survey, 2010-09-11 (the date convention is YYYY-MM-DD), as it was observed in a different IF configuration. We also exclude data from 2012-09-03 when the paddle wheel was broken, resulting in a meaningless measurement of \Tsys (9999 K) in Table~\ref{table:fit_parameters_long} and uncalibrated amplitudes.

\onecolumn

\input{source_info_table.tex}

\clearpage
\onecolumn
\input{fit_parameters_table.tex}

\twocolumn

\end{document}

%% file: gainparameters_table.tex
\begin{table}[t]
\small
\begin{center}
\caption{Gain Curves}\label{table:gaincurves}
\begin{tabular}{lcccc}
\hline 
Line & $\sigma_i^a$ 	     &  $\beta$ 		& $\delta$ & $\sigma_f^b$\\
	&      [K]             & 	  [Hr$^{-1}$]  & [deg$^{-1}$]   &  [K]	\\
\hline 

\nthp	 & 	0.38(2)	 & 	0.024(9)	 & 	0.016(2)	 & 	0.26(2)\\
\hnc	 & 	0.34(1)	 & 	0.028(6)	 & 	0.010(1)	 & 	0.24(1)\\
\hcop	 & 	0.43(2)	 & 	0.028(6)	 & 	0.010(1)	 & 	0.36(2)\\
\hcn	 & 	0.29(2)	 & 	0.040(7)	 & 	0.012(2)	 & 	0.28(2)\\

\hline
\multicolumn{5}{p{7cm}}{{$^a$Absolute flux uncertainty before accounting for systematic gain variations.}}\\
\multicolumn{5}{p{7cm}}{{$^b$Absolute flux uncertainty after accounting for systematic gain variations.}}\\
\end{tabular}
\medskip\\
\end{center}

\end{table}

%% file: source_info_table.tex
\begin{center}
{\footnotesize
\setlength\tabcolsep{1.4ex}
\begin{longtable}{llrrrrcrc}
\caption[Properties]{Properties of PSW Observations of G301}
\label{table:source_info_long} 
\\

\hline \hline \\[-2ex]
   \multicolumn{1}{c}{\textbf{No.$^a$}} &
   \multicolumn{1}{c}{\textbf{Name}} &
   \multicolumn{1}{c}{\textbf{T$_\textrm{sys}$}} &
   \multicolumn{1}{c}{\textbf{El.}} &
   \multicolumn{1}{c}{\textbf{Az.}} &
   \multicolumn{1}{c}{\textbf{MJD}} &
   \multicolumn{1}{c}{\textbf{Time of Day$^b$}} &
   \multicolumn{1}{c}{\textbf{Temp}} &
   \multicolumn{1}{c}{\textbf{Time since May 1st}} \\

   \multicolumn{1}{c}{} &
   \multicolumn{1}{c}{} &
   \multicolumn{1}{c}{\textbf{[K]}} &
   \multicolumn{1}{c}{\textbf{[deg]}} &
   \multicolumn{1}{c}{\textbf{[deg]}} &
   \multicolumn{1}{c}{\textbf{[days]}} &
   \multicolumn{1}{c}{\textbf{[hours]}} &
   \multicolumn{1}{c}{\textbf{[$^{\circ}$C]}} &
   \multicolumn{1}{c}{\textbf{[$10^7$ sec]}} 

   \\[0.5ex] \hline
   \\[-1.8ex]
\endfirsthead

\multicolumn{4}{c}{{\tablename} \thetable{} -- Continued} \\[0.5ex]
  \hline \hline \\[-2ex]
   \multicolumn{1}{c}{\textbf{No.$^a$}} &
   \multicolumn{1}{c}{\textbf{Name}} &
   \multicolumn{1}{c}{\textbf{T$_\textrm{sys}$}} &
   \multicolumn{1}{c}{\textbf{El.}} &
   \multicolumn{1}{c}{\textbf{Az.}} &
   \multicolumn{1}{c}{\textbf{MJD}} &
   \multicolumn{1}{c}{\textbf{Time of Day$^b$}} &
   \multicolumn{1}{c}{\textbf{Temp}} &
   \multicolumn{1}{c}{\textbf{Time since May 1st}}\\

   \multicolumn{1}{c}{} &
   \multicolumn{1}{c}{} &
   \multicolumn{1}{c}{\textbf{[K]}} &
   \multicolumn{1}{c}{\textbf{[deg]}} &
   \multicolumn{1}{c}{\textbf{[deg]}} &
   \multicolumn{1}{c}{\textbf{[days]}} &
   \multicolumn{1}{c}{\textbf{[hours]}} &
   \multicolumn{1}{c}{\textbf{[$^{\circ}$C]}} &
   \multicolumn{1}{c}{\textbf{[$10^7$ sec]}} 

    \\[0.5ex] \hline
   \\[-1.8ex]

\endhead

\hline
\multicolumn{9}{l}{{Continued on Next Page\ldots}} \\
\endfoot

  \\[-1.8ex] \hline \hline
\endlastfoot

0	 & 	2010-07-12\_1	 & 	432	 & 	40.08	 & 	146.79	 & 	55389.297	 & 	7.11	 & 	11.9	 & 	0.62321\\
1	 & 	2010-07-12\_2	 & 	521	 & 	36.40	 & 	146.43	 & 	55389.276	 & 	6.59	 & 	11.6	 & 	0.62302\\
2	 & 	2010-07-12\_3	 & 	496	 & 	37.44	 & 	146.47	 & 	55389.282	 & 	6.74	 & 	11.6	 & 	0.62307\\
3	 & 	2010-07-12\_4	 & 	438	 & 	39.07	 & 	146.63	 & 	55389.291	 & 	6.96	 & 	11.9	 & 	0.62316\\
4	 & 	2010-07-13\_1	 & 	303	 & 	39.18	 & 	146.65	 & 	55390.289	 & 	6.92	 & 	13.3	 & 	0.63178\\
5	 & 	2010-07-14\_1	 & 	193	 & 	34.01	 & 	146.49	 & 	55391.256	 & 	6.13	 & 	9.5	 & 	0.64013\\
6	 & 	2010-07-15\_1	 & 	184	 & 	35.57	 & 	146.42	 & 	55392.263	 & 	6.28	 & 	7.9	 & 	0.64883\\
7	 & 	2010-07-16\_1	 & 	170	 & 	35.43	 & 	146.43	 & 	55393.259	 & 	6.21	 & 	9.3	 & 	0.65744\\
\end{longtable}}
\end{center}

%% file: fit_parameters_table.tex
\begin{center}
{\footnotesize
\setlength\tabcolsep{1.4ex}
\begin{longtable}{lllllllllcc}
\caption[Fits]{Fit Results for PSW Observations of G301}
\label{table:fit_parameters_long} \\

\hline \hline \\[-2ex]

   \multicolumn{1}{c}{} &
   \multicolumn{8}{c}{\textbf{Parameters of Gaussian Fits to Transitions}$^a$} &
   \multicolumn{2}{c}{} \\

   \multicolumn{1}{c}{} &
   \multicolumn{2}{c}{\textbf{\nthp}} &
   \multicolumn{2}{c}{\textbf{\hnc}} &
   \multicolumn{2}{c}{\textbf{\hcop}} &
   \multicolumn{2}{c}{\textbf{\hcn}} &
   \multicolumn{2}{c}{\textbf{Best-fit Locations$^b$}} \\
   
   \multicolumn{1}{c}{\textbf{No.}} &
   \multicolumn{1}{c}{$\mathbf{<a>}$} &
   \multicolumn{1}{c}{$\mathbf{<v>}$} &
   \multicolumn{1}{c}{$\mathbf{<a>}$} &
   \multicolumn{1}{c}{$\mathbf{<v>}$} &
   \multicolumn{1}{c}{$\mathbf{<a>}$} &
   \multicolumn{1}{c}{$\mathbf{<v>}$} &
   \multicolumn{1}{c}{$\mathbf{<a>}$} &
   \multicolumn{1}{c}{$\mathbf{<v>}$} &
   \multicolumn{1}{c}{\textbf{l}} &
   \multicolumn{1}{c}{\textbf{b}} \\
   \multicolumn{1}{c}{} &
   \multicolumn{1}{c}{\textbf{[K]}} &
   \multicolumn{1}{c}{\textbf{[\kms]}} &
   \multicolumn{1}{c}{\textbf{[K]}} &
   \multicolumn{1}{c}{\textbf{[\kms]}} &
   \multicolumn{1}{c}{\textbf{[K]}} &
   \multicolumn{1}{c}{\textbf{[\kms]}} &
   \multicolumn{1}{c}{\textbf{[K]}} &
   \multicolumn{1}{c}{\textbf{[\kms]}} &
   \multicolumn{1}{c}{[deg]} &
   \multicolumn{1}{c}{[deg]}   \\[0.5ex] \hline
   
   \\[-1.8ex]
   \\[-1.8ex]
\endfirsthead

\multicolumn{4}{c}{{\tablename} \thetable{} -- Continued} \\[0.5ex]
  \hline \hline \\[-2ex]

   \multicolumn{1}{c}{} &
   \multicolumn{8}{c}{\textbf{Parameters of Gaussian Fits to Transitions}$^a$} &
   \multicolumn{2}{c}{} \\

   \multicolumn{1}{c}{} &
   \multicolumn{2}{c}{\textbf{\nthp}} &
   \multicolumn{2}{c}{\textbf{\hnc}} &
   \multicolumn{2}{c}{\textbf{\hcop}} &
   \multicolumn{2}{c}{\textbf{\hcn}} &
   \multicolumn{2}{c}{\textbf{Best-fit Locations$^b$}} \\
   
   \multicolumn{1}{c}{\textbf{No.}} &
   \multicolumn{1}{c}{$\mathbf{<a>}$} &
   \multicolumn{1}{c}{$\mathbf{<v>}$} &
   \multicolumn{1}{c}{$\mathbf{<a>}$} &
   \multicolumn{1}{c}{$\mathbf{<v>}$} &
   \multicolumn{1}{c}{$\mathbf{<a>}$} &
   \multicolumn{1}{c}{$\mathbf{<v>}$} &
   \multicolumn{1}{c}{$\mathbf{<a>}$} &
   \multicolumn{1}{c}{$\mathbf{<v>}$} &
   \multicolumn{1}{c}{\textbf{l}} &
   \multicolumn{1}{c}{\textbf{b}} \\
   \multicolumn{1}{c}{} &
   \multicolumn{1}{c}{\textbf{[K]}} &
   \multicolumn{1}{c}{\textbf{[\kms]}} &
   \multicolumn{1}{c}{\textbf{[K]}} &
   \multicolumn{1}{c}{\textbf{[\kms]}} &
   \multicolumn{1}{c}{\textbf{[K]}} &
   \multicolumn{1}{c}{\textbf{[\kms]}} &
   \multicolumn{1}{c}{\textbf{[K]}} &
   \multicolumn{1}{c}{\textbf{[\kms]}} &
   \multicolumn{1}{c}{[deg]} &
   \multicolumn{1}{c}{[deg]}   \\[0.5ex] \hline
  \\[-1.8ex]
\endhead

\hline
\multicolumn{11}{l}{{Continued on Next Page\ldots}} \\
\endfoot

  \\[-1.8ex] \hline \hline
\endlastfoot

4	 & 	1.07(5)	 & 	-42.53(6)	 & 	2.01(4)	 & 	-42.80(4)	 & 	3.32(4)	 & 	-42.70(3)	 & 	2.46(6)	 & 	-43.01(6)	 & 	300.9678	 & 	+1.1446\\
5	 & 	0.87(3)	 & 	-42.49(5)	 & 	1.60(3)	 & 	-42.79(3)	 & 	2.63(3)	 & 	-42.68(2)	 & 	2.12(3)	 & 	-42.98(4)	 & 	300.9672	 & 	+1.1440\\
6	 & 	1.01(3)	 & 	-42.41(4)	 & 	1.78(3)	 & 	-42.74(3)	 & 	2.87(3)	 & 	-42.62(2)	 & 	2.27(3)	 & 	-42.91(4)	 & 	300.9753	 & 	+1.1446\\
7	 & 	1.16(3)	 & 	-42.37(3)	 & 	1.87(2)	 & 	-42.65(2)	 & 	2.99(2)	 & 	-42.59(2)	 & 	2.35(3)	 & 	-42.85(3)	 & 	300.9772	 & 	+1.1440\\
8	 & 	1.98(3)	 & 	-42.53(2)	 & 	2.57(3)	 & 	-42.75(2)	 & 	3.93(3)	 & 	-42.68(1)	 & 	2.84(3)	 & 	-42.83(2)	 & 	300.9659	 & 	+1.1434\\
9	 & 	1.65(3)	 & 	-42.51(2)	 & 	2.36(2)	 & 	-42.77(2)	 & 	3.82(3)	 & 	-42.68(1)	 & 	2.64(3)	 & 	-42.97(3)	 & 	300.9678	 & 	+1.1440\\
10	 & 	1.91(3)	 & 	-42.60(2)	 & 	2.53(3)	 & 	-42.76(2)	 & 	3.94(3)	 & 	-42.69(1)	 & 	2.81(3)	 & 	-42.84(2)	 & 	300.9666	 & 	+1.1434\\
12	 & 	1.08(3)	 & 	-42.39(4)	 & 	1.97(3)	 & 	-42.79(2)	 & 	3.27(3)	 & 	-42.69(2)	 & 	2.46(3)	 & 	-43.05(3)	 & 	300.9722	 & 	+1.1465\\
\end{longtable}}
\end{center}